\documentclass[12pt]{article}

\advance\voffset by -2.0cm \advance\hoffset by -1.25cm
\usepackage[colorlinks,linkcolor=.]{hyperref}

\usepackage{cite}
\usepackage{amsmath}
\usepackage{amssymb}
\usepackage{amsthm}

\usepackage{lscape}
\usepackage{booktabs}
\usepackage{tabularx}
\usepackage{longtable}
\usepackage{ltablex}
\usepackage{caption}
\usepackage{epigraph}

\usepackage{authblk}

\usepackage{multirow}

\usepackage[table]{xcolor}

\newcommand{\RR}{\mathbb{R}}

\newcommand{\ZZ}{\mathbb{Z}}

\newcommand{\half}{{\frac{1}{2}}}

\newcommand{\M}{\mathcal{M}}
\newcommand{\N}{\mathcal{N}}

\newcommand{\Z}{\mathcal{Z}}

\usepackage{adjustbox}

\DeclareMathOperator{\Tr}{Tr}

\DeclareMathOperator{\Weyl}{Weyl}

\newcommand{\avg}[1]{\left\langle#1\right\rangle}

\newcommand{\parens}[1]{\!\left(#1\right)}

\newcommand{\brackets}[1]{\!\left[#1\right]}
\DeclareRobustCommand{\beginProtected}[1]{\begin{#1}}
\DeclareRobustCommand{\endProtected}[1]{\end{#1}}
\newcommand{\piecewise}[4]{\left\{\beginProtected{array}{rl}#1&:#2\\#3&:#4\endProtected{array}\right.}
\newcommand{\column}[2]{
\begin{pmatrix}
#1 \\
#2
\end{pmatrix}}

\newcommand{\twoMatrix}[4]{
\begin{pmatrix}
#1 & #2 \\
#3 & #4
\end{pmatrix}}

\numberwithin{equation}{section}

\def\be{\begin{equation}}
\def\ee{\end{equation}}

\allowdisplaybreaks[3]
\title{Umbral Moonshine and String Duality}
\author[1]{Max Zimet}
\affil[1]{Stanford Institute for Theoretical Physics,

Stanford University, Stanford, CA 94305 USA}
\date{}
\newcommand{\finishRow}[1]{\multicolumn{\the\numexpr 14-#1\relax}{C}{}}

\begin{document}
\maketitle

\begin{abstract}
By studying 2d string compactifications with half-maximal supersymmetry in a variety of duality frames, we find a natural physical setting for understanding Umbral moonshine. Near points in moduli space with enhanced gauge symmetry, we find that the Umbral symmetry groups arise as symmetries of the theory. In one duality frame -- a flux compactification on $T^4/Z_2\times T^4$ -- the 24-dimensional permutation representations of the Umbral groups act on D1-branes strung between a set of NS5-branes. The presence of these NS5-branes is used to explain the Umbral moonshine decompositions of the K3 twining genera, and in particular of the K3 elliptic genus. The fundamental string in this frame is dual to the type IIA string on K3$\times T^4$ and to a compactified heterotic little string theory. The latter provides an interesting example of a little string theory, as the string-scale geometry transverse to the 5-brane plays an important role in its construction.
\end{abstract}

\pagebreak
\tableofcontents

\hypersetup{linkcolor=blue}

\section{Introduction}
Compactifications of string theory with half-maximal supersymmetry have played a central role in physics. For example, studies of such compactifications have yielded exciting developments in K3 mirror symmetry\cite{Aspinwall:1994rg}, other string dualities\cite{hull:unity,w:various}, and black hole microstate counting\cite{StromingerVafa,DVV}. More recently, it was argued that compactifications of this type to low dimensions (and orbifolds thereof) can be used to explain a number of interesting aspects of Monstrous moonshine\cite{Conway:1979kx,FLM,PPV,PPV2}. In this paper, we use such compactifications to address another interesting problem in mathematical physics: Umbral moonshine\cite{Eguchi:2010ej,Cheng:2012tq,Cheng:2013wca,Cheng:2014zpa}.

The subject of Umbral moonshine began when \cite{Eguchi:2010ej} noticed that the elliptic genus of K3, $Z^{K3}$ -- a supersymmetric index which is invariant under marginal deformations -- decomposes naturally into two pieces -- which we will call the `singular part' and the `Umbral part,' for reasons that will become clear later -- where the latter is constructed from a mock modular form whose coefficients are dimensions of representations of the Mathieu group $M_{24}$, one of the 26 sporadic simple groups. This result was then generalized to include 22 other similar decompositions of the K3 elliptic genus. First, a purely number theoretic construction was found that associates a vector-valued `Umbral mock modular form' $H^X$ to each of the 23 Niemeier lattices (the positive-definite even self-dual lattices in 24 dimensions with roots)\cite{Cheng:2013wca}. The Niemeier lattices, $L^X$, are uniquely specified by their root systems, $X$, each of which is of ADE type. As we review below, the groups whose representations' dimensions appear in these Umbral mock modular forms -- which we call the `Umbral groups,' $G^X$ -- are certain symmetry groups of their corresponding Niemeier lattice; in particular, $M_{24}$ is associated to the $X=A_1^{24}$ Niemeier lattice.\footnote{The decomposition of $Z^{K3}$ associated to $A_1^{24}$ is related to the $\N=4$ superconformal character decomposition.} Then, \cite{Cheng:2014zpa} related this back to string theory on K3 by introducing a (non-invertible) transformation that maps the functions $H^X$ to new functions, $Z^{X,U}$, and demonstrating that $Z^{K3}=Z^{X,S}+Z^{X,U}$, where $Z^{X,S}$ is the elliptic genus of the ALE space\footnote{ALE (asymptotically locally Euclidean) spaces are hyper-K\"ahler 4d gravitational instantons that asymptote (sufficiently rapidly) to $\RR^4/\Gamma$ at infinity for some finite group $\Gamma$ (which turns out to always be a subgroup of $SU(2)$). These have an ADE classification. The $A_n$ cases are the multi-center Eguchi-Hanson geometries ($A_0$ is $\RR^4$ and $A_1$ is the Eguchi-Hanson manifold). Also of interest in this paper will be ALF (asymptotically locally flat) gravitational instantons, which asymptote to $(\RR^3\times S^1)/\Gamma$. These have a related classification \cite{minerbe:mass,minerbe:ALF,minerbe:rigidTaubNUT,minerbe:Dk,chen:ALF}, which in most cases corresponds to that of the ALE spaces via decompactification. The $A_n$ cases are the multi-Taub-NUT geometries ($A_{-1}$ is $\RR^3\times S^1$ and $A_0$ is the Taub-NUT space), while the $D_n$ cases may in a sense be obtained by combining the A-type ALF manifolds with the Atiyah-Hitchin $D_0$ manifold, as described in \cite{sen:d6KK}. The $D_n$ ALF manifolds were explicitly constructed in \cite{cherkis:dkGravInst,cherkis:singMonopolesDk,cherkis:dkInst}. We note that there are no E-type ALF manifolds. Explicitly, the ALE and ALF classifications respectively yield $A_k,k\ge 0$, $D_k,k\ge 4$, and $E_6,E_7,E_8$; and $A_k,k\ge-1$ and $D_k,k\ge0$.} with the corresponding ADE singularity at its center. Further evidence that Umbral moonshine is related to K3 string theory was provided by \cite{Cheng:2010pq,Gaberdiel:2010ch,Gaberdiel:2010ca,Eguchi:2010fg,Sen:2010ts,Creutzig:2013mqa,cheng2015landau,slowpaper,mz:walls}, where the twining genera $Z_g^{K3}$ (generalizations of the elliptic genus that contain information about the actions of symmetries on BPS states) corresponding to all ($\N=(4,4)$-preserving) symmetries $g$ of K3 non-linear sigma models (NLSMs) were classified and shown to be equal to the twining genera of \cite{Cheng:2014zpa}, which were again constructed by applying a transformation to mock modular forms $H_g^X$ defined in \cite{Cheng:2013wca}.\footnote{A minor correction to the prescription for obtaining $H_g^X$ was provided in \cite{Cheng:rademacher}.}\textsuperscript{,}\footnote{More precisely, it was shown that all K3 NLSM twining genera appear in the list of functions proposed in \cite{Cheng:2014zpa}. The converse, while extremely well-motivated and known to be true in most cases, has not yet been completely proven.}

In Monstrous moonshine, similar observations (where we replace the Umbral mock modular forms by the $j$-function of number theory, their twined variants by the McKay-Thompson series, and the Umbral groups by the Monster group) led to the construction of a 2d CFT \cite{FLM} with the Monster group as its symmetry group. It was thus natural to hope that there were points in the moduli space of string theories on K3 that had these Umbral groups as symmetries. Unfortunately, this hope was dashed by \cite{Gaberdiel:2011fg}, which showed that there is no K3 NLSM with $M_{24}$ symmetry. In retrospect (i.e., now that we have the formulae of \cite{Cheng:2014zpa}), since K3 surfaces can develop at most 19 singularities (or 20, if we include the B-field of string theory)\cite{w:various}, it would be somewhat surprising if this hope were realized, since the ADE root systems associated to the Niemeier lattices all have rank 24. We are thus led to the following interesting problem: for each Umbral group $G^X$, construct a supersymmetric sigma model with $G^X$ symmetry whose target space has the appropriate rank 24 singularity and for which a supersymmetric index may be computed that implies the decomposition $Z^{K3}=Z^{X,S}+Z^{X,U}$. While this problem has not yet been solved, we note that infinite-dimensional modules whose graded characters equal the Umbral mock modular forms have been proven to exist\cite{Gannon:2012ck,Duncan:Umbral}, and that such modules were even explicitly constructed in the cases of $E_8^3$\cite{Harvey:E8}, $A_6^4$ and $A_{12}^2$ \cite{duncan:someAs}, $D_6^4$, $D_8^3$, $D_{12}^2$, and $D_{24}$ \cite{cheng:someDs}, and $D_4^6$ \cite{cheng:d46}.

In this paper, we solve this problem, except instead of ADE singularities our 8-dimensional target spaces will have the T-dual NS5-brane throats with torsion. (Because of this duality \cite{hull:unity,ooguri:ADE,w:comments,s:coulomb,brodie:2dMirror,sethi:mirror,tong:mirror,harvey:monopoles3,jensen:doubledKK}, they have the same elliptic genera.) The decomposition of the K3 elliptic genus into Umbral and singular parts arises naturally in the study of these spaces. Global considerations from string dualities ensure that their (modified) elliptic genera are related to the elliptic genus of K3, while local considerations imply the existence of the appropriate ADE throats.

\begin{figure}
\includegraphics[width=\textwidth]{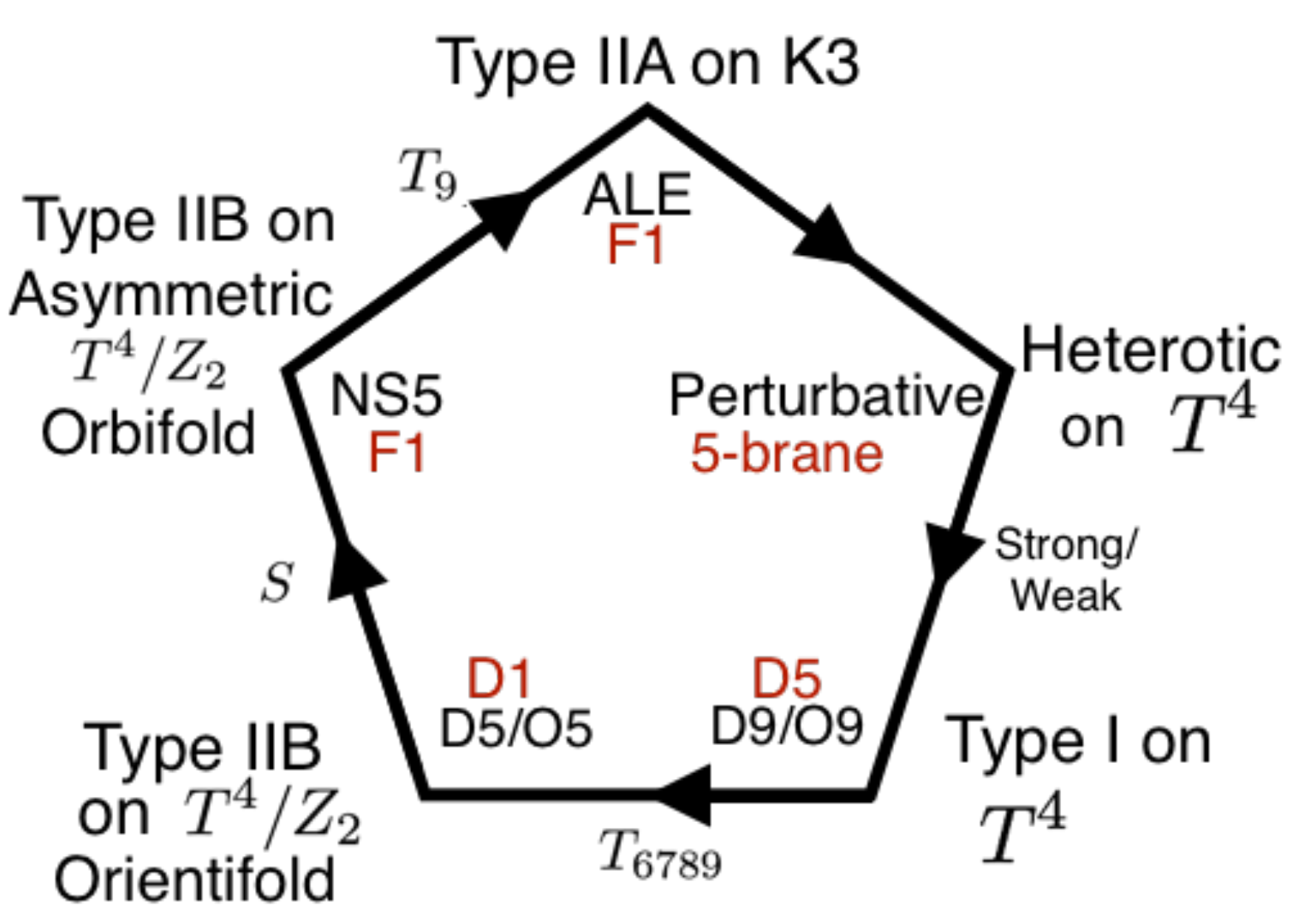}
\caption{The different duality frames that play a role in this paper. The inside of the pentagon labels the mechanism for gauge symmetry enhancement in black and the dual of the type IIA fundamental string in red. In the upper left frame, the $Z_2$ fixed points play a role in the development of enhanced gauge symmetry that is analogous to that of O5-planes in the bottom left frame. The 5-branes of the two frames on the right wrap their respective $T^4$. $T_{\mu_1\ldots\mu_k}$ refers to T-duality in the directions $\mu_1,\ldots,\mu_k$.}\label{fig:dualityFrames}
\end{figure}

We describe these string backgrounds by studying a variety of different duality frames, as summarized in figure \ref{fig:dualityFrames}. We begin with type IIA on K3. This is dual to both heterotic and type I on $T^4$ (since, after all, $SO(32)$ heterotic and type I string theory are already related by strong-weak duality in 10 dimensions). One of the most interesting aspects of these dualities is that we are frequently able to map BPS excitations across them, since supersymmetry lets us study BPS states even at strong coupling. For example, the perturbative heterotic string is mapped to the type I D1-brane\cite{w:evidence,dab:hetSol,hull:stringString} and the type IIA NS5-brane wrapping K3\cite{Harvey:soliton,sen:soliton}. More importantly, for the purposes of this paper, is the mapping of the type IIA fundamental string: in the heterotic frame it is a $T^4$-wrapped 5-brane, which in type I maps to a $T^4$-wrapped D5-brane\cite{w:smallInst,Harvey:soliton,sen:soliton}. The Higgs branch of the worldvolume theories of these last two solitons is the ADHM moduli space of an instanton \cite{w:smallInst,strominger:hetSol,Harvey:instanton,douglas:instanton,adhm}. However, we will be more interested in the Coulomb branch, as this describes the degrees of freedom which arise from the type IIA string probing the K3 of its target space\cite{w:smallInst,s:K3}.

The roles of these various duality frames in relation to Umbral moonshine are as follows. We first compactify on a 4-torus to two dimensions. The K3 elliptic genus will arise as part of a modified elliptic genus for the type IIA fundamental string. The heterotic frame makes it clear that we can have an ADE gauge symmetry with rank up to 24. In particular, this frame makes it clear that we can have gauge groups that are not subgroups of $SO(32)$; in the other duality frames, this requires a non-perturbative explanation. (In the original type IIA frame, any non-Abelian gauge symmetry requires a non-perturbative explanation. For example, in six dimensions either an ADE singularity develops in the K3 surface, or the background is smooth, but the metric and B-field are tuned in such a way that the K3 sigma model is still singular \cite{w:various,townsend:gaugeSymm,aspinwall:theta,w:comments,Aspinwall:1996mn}.) It also allows us to confront the pesky issue of anomalies in two dimensions: in the $g_s\to 0$ limit, where we can argue that anomalies do not plague the consistency of our setup, the heterotic little string theory (HLST) on the 5-brane corresponding to the IIA fundamental string remains interacting. Finally, in this frame it is easy to see how the Umbral symmetry groups arise as symmetries of string theory. Namely, the Umbral groups arise, as in \cite{kachru20163d}, as the symmetry groups that remain unbroken as we perturb infinitessimally away from points with rank 24 non-Abelian gauge symmetry in order to break the gauge symmetry to its maximal Abelian subgroup.\footnote{\cite{kachru20163d} worked with 3d string compactifications, whereas we find it necessary to go down to two dimensions. The reason for this difference is that we want to be able to vary the string coupling constant while remaining at a Niemeier point.}

Next, we argue that the singular pieces of the elliptic genus correspond to singularities in the moduli space of this compactified little string theory. Since the global aspects of the moduli space are unimportant for studying these singularities, we pass to the IR gauge theory, which is most naturally studied in the type I frame and its T-duals.

Finally, we act with S-duality and find a frame where the target space of the fundamental string is the moduli space of the heterotic little string theory. In particular, at points of enhanced gauge symmetry the singularities of the HLST moduli space correspond to NS5-branes in this frame. Gauge bosons arise from D1-branes strung between these NS5-branes. This frame has no R-R backgrounds turned on, so we are in a similar situation to that of \cite{mcGreevy:nonGeo}: we have a flux compactification that we can study via the non-linear sigma model of the fundamental string. This too allows us to confront the anomaly by passing from string theory to a 2d NLSM. Lastly, we observe that the NS5-branes in this frame are particularly interesting from the point of view of moonshine, as those of the aforementioned D1-branes which correspond to a set of simple roots for the corresponding $L^X$ transform in the natural permutation representations of the Umbral groups described in \cite{Cheng:2013wca}.

As the final arrow in figure \ref{fig:dualityFrames} illustrates, if our purpose was simply to reach this final duality frame, we could have simply started in IIA on $T^4/Z_2$ and T-dualized one of the compact dimensions \cite{vafa:dManifolds,kutasov:orbSolitons,Sen:1996na}. This yields an asymmetric orbifold. (This also follows from the other way of arriving at this duality frame, by performing S-duality on an orientifold, since S-duality maps worldsheet orientation reversal to $(-1)^{F_L}$ \cite{kutasov:orbSolitons}.) At general points in moduli space, where the asymmetric orbifold description no longer applies, NS5-branes emerge from the $Z_2$ fixed points and NS-NS backgrounds are turned on, as charges are not canceled locally. We again arrive at the picture described in the preceding paragraphs. However, because many ideas are highlighted more clearly in the other duality frames, as explained above, we will take the long road around figure \ref{fig:dualityFrames}.

We now want to relate this discussion back to the worldsheet NLSM of the dual fundamental type IIA string on K3$\times T^4$. We can only compare supersymmetric indices which are invariant under marginal deformations, since the perturbative points in string moduli space where these worldvolume theories are valid differ. However, this still leaves us with plenty of interesting ways with which we can test this duality. In particular, the (modified) elliptic genera of these theories should agree.

We thus arrive at our connection with mock Jacobi forms: when we compactify to two dimensions, there are certain points in moduli space where the theory develops an enhanced non-Abelian gauge symmetry of rank 24, and near these points the (modified) elliptic genera of these NLSMs decompose naturally according to the formulae of \cite{Cheng:2014zpa}. This decomposition has a natural interpretation (up to certain caveats discussed in \S\ref{sec:duality}): we are gluing together two non-compact spaces in order to form a compact target space. This is in accordance with the understanding that mock automorphic forms arise naturally in string theory on non-compact spaces\cite{w:mockVafa,troost:mock,eguchi:mock,troost:mockMeasure,Dabholkar:2012nd,murthy:mockGLSM,troost:mockGLSM}. Finally, our picture essentially explains the correspondence between the twined functions $Z^{X,S}_g$ and $Z^{X,U}_g$ of Umbral moonshine and the twining genera $Z^{K3}_g$ of K3 NLSMs.

The outline of the rest of this paper is as follows. In section \ref{sec:reprise}, we review the pertinent aspects of Umbral moonshine, as well as the explanation in \cite{kachru20163d} of the physical role of the Umbral groups, which we modify slightly. Section \ref{sec:duality} is devoted to a study of the above dualities. In section \ref{sec:ant}, we observe that our constructions do not appear to rely on the non-Abelian gauge group (being slightly Higgsed away from) having rank 24, and we are thus led to suggest the possibility of a generalization of Umbral moonshine to an Antumbral moonshine.\footnote{Umbral moonshine being a subclass of the Antumbral variety assumes that the antumbra is defined to include the umbra. We are unaware of any conventions regarding this point, but we note that NASA's Navigation and Ancillary Information Facility defines the penumbra to include the umbra \cite{nasa:penumbra}, so our definition seems sensible.} We conclude in section \ref{sec:conclude}. Because of the importance of the non-perturbative appearance of extra NS5-branes in our story, in Appendix \ref{sec:fieldReview} we review many aspects of field theories with 8 supercharges and their role in probing these phenomena in string theory.

\section{Moonshine Reprise}\label{sec:reprise}

\subsection{Decompositions of the K3 Elliptic Genus}

We begin with a brief review of some aspects of Umbral moonshine; the reader is referred to appendix \ref{a:basics} for mathematical background and definitions of special functions. The elliptic genus of a $(4,4)$ superconformal field theory (SCFT) whose holomorphic and anti-holomorphic central charges are equal and given by $c$ is\footnote{The elliptic genus can be defined for many other theories, but this is the only case of relevance for this paper.}
\be Z(\tau,z) = \Tr_{RR}\parens{(-1)^{J_0+\bar J_0}q^{L_0}\bar q^{\bar L_0}y^{J_0}}\qquad q=e^{2\pi i \tau},y=e^{2\pi i z}\ .\label{eq:eg}\ee
The $RR$ subscript indicates that the trace is restricted to the Ramond-Ramond sector; $L_0,\bar L_0$ are Virasoro generators (normalized so that they vanish in RR ground states); and $J_0,\bar J_0$ are the Cartan generators of the left- and right-moving $SU(2)$ R-symmetries of the $\N=4$ superconformal algebra. This trace only receives contributions from states that are BPS with respect to the right-moving algebra; these states have $\bar L_0=0$, so the elliptic genus is actually holomorphic in both $\tau$ and $z$. In fact, it is a weak Jacobi form of weight 0 and index $c/6$. Setting $y=1$ yields the Witten index, which for NLSMs is the Euler characteristic of the target space. Finally, we note that the elliptic genus is invariant under supersymmetry-preserving marginal deformations of the SCFT.

Of particular interest is the $(4,4)$ NLSM relevant for studying perturbative type IIA on K3. This has an 80-dimensional moduli space\cite{Aspinwall:1994rg}
\be\label{eq:k3moduli} \M_{K3}=O^+(\Gamma^{20,4})\backslash O^+(20,4)/(O(20)\times SO(4))\ ,\ee
which parametrizes the metric and B-field on the K3 surface.\footnote{$\Gamma^{20,4}$ is the unique even unimodular lattice of signature $(20,4)$. We denote by $O^+(20,4)$ the subgroup of $O(20,4)$ whose maximal compact subgroup is $O(20)\times SO(4)$. $O(\Gamma)$ refers to the automorphism group of a lattice $\Gamma$, and $O^+(\Gamma^{20,4})$ is the subgroup of $O(\Gamma^{20,4})$ contained in $O^+(20,4)$.}\textsuperscript{,}\footnote{See \cite{Nahm:1999ps,slowpaper} for the reason for the pluses.} In particular, there is a $T^4/Z_2$ orbifold point, at which the elliptic genus may be easily computed:
\be\label{eq:k3eg} Z^{K3}(\tau,z)=2\chi_{0,1}=2y+2y^{-1}+20+O(q)\ .\ee
Since $\M_{K3}$ is connected, all K3 NLSMs have the same elliptic genus.

The elliptic genus of K3 has an Umbral decomposition corresponding to each of the 23 Niemeier lattices, $L^X$. These are the only even unimodular lattices of signature $(24,0)$ that have roots, and they are uniquely specified by their ADE-type (i.e. union of simply laced) root systems, which we denote by $X$. To each such lattice, we associate the Umbral group
\be \label{eq:umbralGp} G^X=O(L^X)/\text{Weyl($X$)}\ ,\ee
where Weyl($X$) is the Weyl group of the root system $X$.\footnote{In general, $X$ will not be irreducible. We define the Weyl group of a reducible root system to be the direct product of the Weyl groups of its irreducible components. We similarly define the Weyl vector of $X$ to be the direct sum of the Weyl vectors of its irreducible components.} We note that the presence of `glue vectors' (see, e.g., \cite{ConwaySloane}) obscures the symmetries of these lattices. For example, $G^{A_1^{24}}$ is not the symmetric group $S_{24}$, but rather the subgroup $M_{24}$.

We can now formulate the Umbral decompositions of $Z^{K3}$:
\be\label{eq:umbralDecomp} Z^{K3}(\tau,z) = Z^{X,S}(\tau,z) + Z^{X,U}(\tau,z)\ ,\ee
where $Z^{X,S}$ is the elliptic genus of the ALE space specified by $X$,\footnote{Or, rather the sum of the elliptic genera of the ALE spaces specified by the irreducible components of $X$.} and $Z^{X,U}$ is the Umbral part of the decomposition. More precisely, there are vector-valued mock modular forms $H^X(\tau)=\parens{H^X_r(\tau)}$ whose coefficients are (for the most part -- see \S2.1 of \cite{Cheng:rademacher} for the precise statement) dimensions of representations of $G^X$ such that if we define the mock Jacobi forms
\be\label{eq:phiX} \phi^X(\tau,z) = \frac{-i\theta_1(\tau,mz)\theta_1(\tau,(m-1)z)}{\eta^3(\tau)\theta_1(\tau,z)}\sum_{r\in \ZZ/2m\ZZ}H_r^X(\tau)\theta_{m,r}(\tau,z)\ ,\ee
then
\be\label{eq:zUmbral} Z^{X,U}(\tau,z) = \frac{1}{2m}\sum_{a,b\in \ZZ/m\ZZ} q^{a^2}y^{2a}\phi^X\parens{\tau,\frac{z+a\tau+b}{m}} \ .\ee
In these equations, $m$ is the Coxeter number of an irreducible component of $X$; for the Niemeier lattices, $m$ is the same for any choice of irreducible component. This implies that the root systems are of the form
\be\label{eq:roots} X=A_{m-1}^{d_A}\oplus D_{m/2+1}^{d_D}\oplus \parens{E^{(m)}}^{d_E}\ ,\ee
where $E^{(m)}=E_6,E_7,E_8$ for $m=12,18,30$. With this observation, we may expand the functions $Z^{X,S}$ as
\be\label{eq:moreSingular} Z^{X,S} = d_A Z^{A_{m-1},S} + d_D Z^{D_{m/2+1},S} + d_E Z^{E^{(m)},S}\ .\ee
Finally, since the ALE spaces are non-compact, sigma models with ALE target spaces have continuous spectra. The functions $Z^{X,S}$ are the holomorphic functions obtained by restricting to the discrete part of these spectra; including the continuous part yields a non-holomorphic harmonic Maass-Jacobi form \cite{BruinierFunke2004,bringmann:maassJacobi,Dabholkar:2012nd}.

\subsection{Physical Origin of the Umbral Groups}\label{sec:origin}

In order to find a physical role for symmetries of Niemeier lattices, and for the Umbral groups in particular, \cite{kachru20163d} studied heterotic string theory on $T^7$. At certain points in the moduli space of this string compactification, the Umbral groups $G^X$ arise as symmetry groups of the theory. This relies upon the result of \cite{sen:strong} that the moduli space of heterotic string theory on $T^7$ is
\be\label{eq:sen3d} O(\Gamma^{24,8})\backslash O(24,8)/(O(24)\times O(8))\ .\ee
At a point in moduli space where $\Gamma^{24,8}$ decomposes as $L^X\oplus E_8(-1)$,\footnote{$E_8(-1)$ refers to the lattice related to the $E_8$ root lattice by negation of the quadratic form.} the symmetries $O(L^X)$ of $L^X$ are also symmetries of the string theory. At these points in moduli space, the theory has a rank 24 non-Abelian gauge symmetry (in addition to some Abelian gauge symmetry). By deforming away in a direction that completely breaks the non-Abelian gauge symmetry to its maximal torus while preserving as much of the symmetry group $O(L^X)$ as possible, we arrive at a point in moduli space with symmetry group $G^X$.\footnote{Actually, even at the point with enhanced gauge symmetry the symmetry group is $G^X$, as the Weyl group consists of gauge symmetries.} In order to be more specific, we let $\lambda$ be the Weyl vector of $X$ and choose orthonormal bases $e_i\in L^X\otimes\RR,i=1,\ldots 24$ and $\tilde e_j\in E_8(-1)\otimes\RR,j=1,\ldots,8$, where $e_1=\lambda/|\lambda|$. The deformation then implements
\be\label{eq:changebasis} \column{p_{L,1}^\text{new}}{p_{R,1}^\text{new}} = \twoMatrix{\cosh\epsilon}{\sinh\epsilon}{\sinh\epsilon}{\cosh\epsilon}\column{p_{L,1}^\text{old}}{p_{R,1}^\text{old}}\ .\ee
Note that $(p_{L,1}^\text{old})^2-(p_{R,1}^\text{old})^2=(p_{L,1}^\text{new})^2-(p_{R,1}^\text{new})^2,$ so this transformation is in $O(24,8)$. It breaks $O(L^X)$ to its maximal subgroup that fixes $\lambda$; this happens to be $G^X$. Any choice of $\lambda$ other than the Weyl vector that completely breaks the non-Abelian gauge symmetry yields a point in moduli space whose symmetry group is a subgroup of $G^X$.

The moduli of \eqref{eq:sen3d} include the dilaton. We will need to be able to vary the dilaton while remaining at (or near) a Niemeier point in moduli space. Therefore, we will study heterotic string theory on $T^8$, where \eqref{eq:sen3d} is the Narain moduli space.\footnote{The full U-duality group is now infinite-dimensional \cite{Sen:1995qk}.} All of the above results carry over seamlessly to this case.

\subsection{Modified Elliptic Genus}\label{sec:mms}

Below, when we compactify down to two dimensions, the elliptic genus will not be well-suited for our purposes, as it will vanish. This is clear at points in the type II moduli space where we have a product $K3\times T^4$, since fermion zero modes from the $T^4$ cause the elliptic genus to vanish. We therefore introduce the modified elliptic genus, or helicity supertrace \cite{Kiritsis:1997hj}. From a worldsheet perspective, this is obtained by inserting $(\bar J_0)^2$ into the trace in \eqref{eq:eg}. In the case of a $T^4$ sigma model, one may argue that it is a deformation-invariant index \cite{mms} if one restricts to states with no charge under the $U(1)^4$ current algebra. This continues to hold for $K3\times T^4$. We again do not count states with extra central charges, so that the index does not jump at non-generic points in moduli space. We can also regard the helicity supertrace as a computation in spacetime, where it computes signed counts of supersymmetry multiplets. To do so, we compactify the final spatial direction on a large circle and regard the string worldsheet as describing a static gauge string wrapping this circle. Then, the helicity supertrace counts multiplets of the $\N=16$ quantum mechanics that arises via dimensional reduction from 2d $\N=(8,8)$. (Analogously, the $T^4$ index can be thought of as counting 5d multiplets \cite{mms}.)

The $T^4$ index can be computed as follows. We begin with the partition function
\begin{align} \label{eq:part} 
\Z(\tau,z) &= \Tr_{RR}\parens{(-1)^{J_0+\bar J_0}q^{L_0}\bar q^{\bar L_0}y^{J_0}\bar y^{\bar J_0}} \nonumber\\
&=\parens{\frac{\theta_1(\tau,z)}{\eta(\tau)}}^2\frac{1}{\eta(\tau)^4}\overline{\parens{\frac{\theta_1(\tau,z)}{\eta(\tau)}}^2\frac{1}{\eta(\tau)^4}} \ .
\end{align}
(We have restricted to states with no winding or momenta, as explained above.)
Then, we insert $(\bar J_0)^2$ by differentiating twice with respect to $\bar y$:
\be \label{eq:mms} \hat Z^{T^4}(\tau, z) = \partial_{\bar y}^2 \Z|_{\bar y=1} = - 2 \parens{\frac{\theta_1(\tau,z)}{\eta(\tau)}}^2\frac{1}{\eta(\tau)^4} \ . \ee
Similar reasoning yields the index for $K3\times T^4$:
\be \label{eq:mms2} \hat Z^{K3\times T^4}(\tau,z) = Z^{K3}(\tau,z)\hat Z^{T^4}(\tau, z)\ .\ee
It is via this formula that the K3 elliptic genus will enter into our story.


\subsection{K3 Twining Genera}\label{sec:twining}

\begin{figure}
\begin{center}
\includegraphics[width=0.5\textwidth]{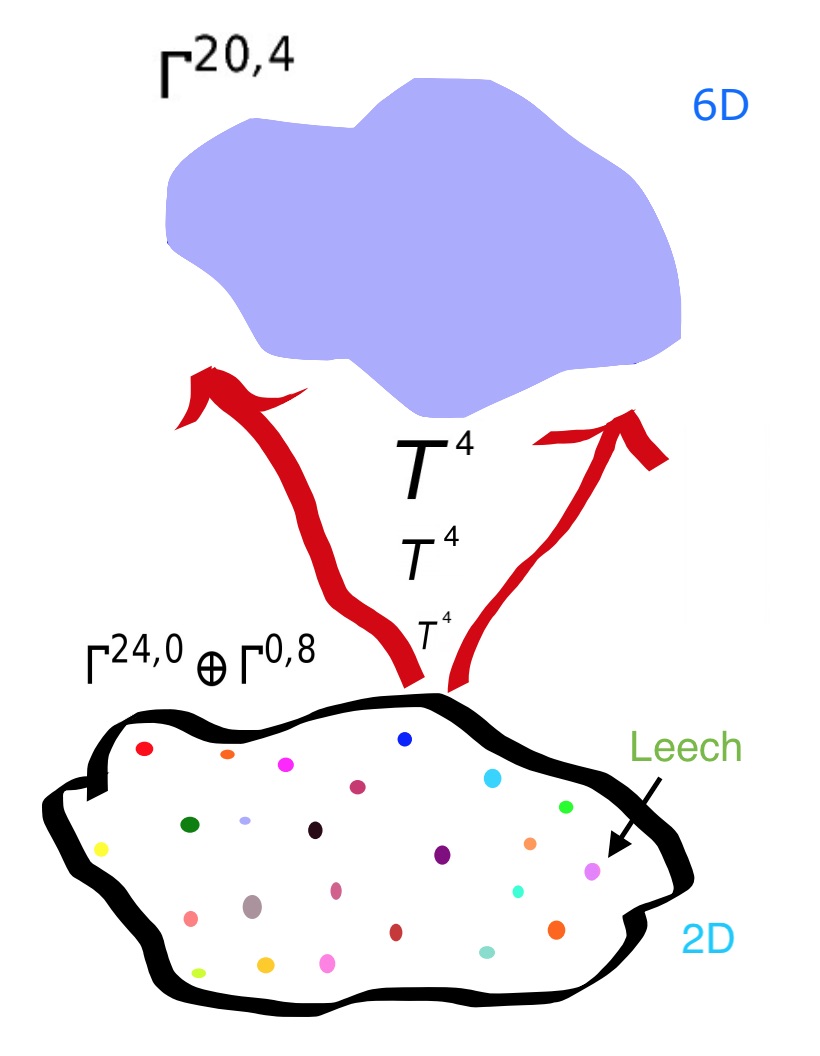}
\caption{Decompactification to six dimensions. Figure borrowed from \cite{kachru20163d} and slightly modified, with permission from the authors. The special points in 2D where $\Gamma^{24,8}$ decomposes as $\Gamma^{24,0}\oplus\Gamma^{0,8}$ are highlighted. The 23 of these corresponding to the Niemeier lattices will play an important role in this paper. The final one, corresponding to the Leech lattice, does not have enhanced gauge symmetry.}\label{fig:decompactify}
\end{center}
\end{figure}

As in \cite{kachru20163d}, we now briefly study Umbral symmetries that survive decompactification to 6 dimensions, as illustrated in figure \ref{fig:decompactify}. Mathematically, they are characterized by the fact that they fix a 4-plane in $L^X\otimes\RR$, while physically they are symmetries of type IIA string theory on K3 that preserve the $\N=(1,1)$ supersymmetry. In fact, subgroups of the Umbral groups consisting of such symmetries are precisely the supersymmetry-preserving discrete symmetry groups that arise at some point in the moduli space \cite{slowpaper}. We restrict to points without enhanced gauge symmetry, where the K3 NLSM is non-singular. Then, these symmetries correspond to symmetries of K3 NLSMs that preserve the $\N=(4,4)$ superconformal algebra and the spectral flow generators, i.e. they fix a negative-definite 4-plane of the space $\Gamma^{20,4}\otimes\RR$ of R-R ground states. To such a symmetry, we can associate a twining genus $Z^{K3}_g$ which is defined by inserting $g$ into the trace \eqref{eq:eg}. Clearly, we can define an analogous index for K3$\times T^4$ by inserting $g (\bar J_0)^2$ instead. We find
\be\label{eq:twine} \hat Z^{K3\times T^4}_g(\tau,z) = Z_g^{K3}(\tau,z)\hat Z^{T^4}(\tau,z) \ . \ee
Therefore, K3 twining genera can easily be studied in the 2d compactifications that interest us.

Just as the elliptic genus may be decomposed into singular and Umbral parts, one can similarly decompose the twining genera \cite{Cheng:2014zpa}. Now, one must take care in choosing $X$, as $g$ may not always arise as an automorphism of $L^X$. Building off of ideas in \cite{nikulin2013kahlerian}, it was explained in \cite{slowpaper} how this should be done.\footnote{Actually, for some symmetries there does not exist any choice of $X$, and the following discussion applies only if we take $L^X$ to be the Leech lattice and $G^X=O(L^X)=Co_0$ to be the `Conway-0' group. Although there is an associated Conway moonshine \cite{duncan:2014eha,duncan:2015xoa}, which also yields K3 twining genera, the relationship between this construction and that of \cite{Cheng:2014zpa} is still mysterious, and we will not consider such symmetries. However, $Co_0$ and the Leech lattice do seem to have quite a lot to do with Umbral moonshine \cite{slowpaper,mz:walls}.} For an appropriate choice of $X$ (which is usually not unique), $g$ may then be thought of as a 4-plane fixing element of $O(L^X)$. The subgroup $\avg{g}\subset O(L^X)$ is isomorphic to a subgroup of $G^X$, so it inherits the 24-dimensional permutation representation of $G^X$ that acts on the simple roots of $X$. By studying this representation, functions $Z^{X,S}_g$ were conjectured in \cite{Cheng:2014zpa} to arise as the twining genera of the NLSM with target space the associated ALE singularity. It was proven that the K3 twining genera decompose as
\be Z_g^{K3}(\tau,z) = Z^{X,S}_g(\tau,z) + Z^{X,U}_g(\tau,z) \ ,\ee
where $Z^{X,U}_g$ is constructed in a manner similar to that of \eqref{eq:phiX}, \eqref{eq:zUmbral} from twined Umbral mock modular forms.

\section{String Duality and Umbral Moonshine}\label{sec:duality}

In this section, we discuss how various dual descriptions of 2d string compactifications shed light on Umbral moonshine. In order for this paper to be self-contained, field theories with 8 supercharges and their role in probing stringy phenomena are reviewed in Appendix \ref{sec:fieldReview}.

\subsection{Duality in 6d}\label{sec:dual6d}

We begin with type IIA string theory on K3. We are interested in following the BPS degrees of freedom of the fundamental string, which are counted by the elliptic genus, as we change to other duality frames. Figure \ref{fig:dualityFrames}, Appendix \ref{sec:fieldReview}, and the references therein explain how this is done. In short, this string becomes a wrapped 5-brane in heterotic string theory on $T^4$, a wrapped D5-brane in type I string theory on $T^4$, and a D1-brane in the type IIB orientifold obtained by T-dualizing the compact dimensions of the type I frame. S-dualizing this last frame again yields a fundamental string, and a T-duality returns us to the original type IIA frame. The elliptic genus of this string's worldvolume theory (minus the center of mass degrees of freedom, which yield an $\RR^4$ factor in the moduli space of the string\footnote{Since we are in 2d, by `moduli space' we really mean the target space of the low-energy NLSM.}\textsuperscript{,}\footnote{We refer to the center of mass moduli as Higgs moduli, since they arise as vevs of scalars in hypermultiplets. However, they are not lifted on the Coulomb branch.}) is the same in each frame. This is also clear in the heterotic and type I frames, where we have a compactified 6d $SU(2)$ gauge theory, whose classical Coulomb branch is $T^4/Z_2$. We may think of the elliptic genus as a signed count of states of these soliton strings wrapped around a large circle \cite{w:smallInst}.

However, we are concerned with points in moduli space where quantum corrections are significant. For, we are interested in the physics of points with enhanced gauge symmetry. On the probe string, this manifests itself via hypermultiplets which become light; integrating these out yields singularities in the Coulomb branch. This is easy to see in the worldvolume of the D1-brane, which hosts a $(4,4)$ gauge theory whose Coulomb branch describes the metric and B-field in the infinite throat transverse to a set of NS5-branes \cite{s:coulomb}. The ADE-type gauge symmetry in spacetime becomes a global symmetry on the D1-brane. In fact, in the S-dual frame where the D1-brane becomes a fundamental string, these NS5-branes exist in spacetime and are responsible for enhanced gauge symmetry. Na\"ively, since these NS5-branes arose from D5-branes in the dual description, which in turn came from D9-branes in the type I frame, we can only have 16 NS5-branes. This is in contradiction with the heterotic frame, where it is known that non-Abelian gauge groups with rank up to 20 are possible. The resolution is that extra NS5-branes come into existence in accordance with duality.

\subsection{Further Compactification}

As discussed in section \ref{sec:origin}, our ultimate goal is heterotic string theory on $T^8$. As an intermediate step, let us compactify a $T^3$ transverse to the HLST; after the chain of dualities we have just described, this corresponds to compactifying a $T^3$ transverse to the D1, so that we are on the IIB orientifold $T^3\times (T^4/Z_2)$. As above, non-perturbative effects can create extra D-branes. Due to the presence of the D1, we refrain from T-dualizing the new $T^3$, so the new D-branes are D5-branes wrapping the $T^3$;\footnote{To see this very explicitly, we can imagine removing the probe brane temporarily. Then, we can follow the same logic as in Appendix \ref{sec:fieldReview} and study non-perturbative gauge symmetry enhancement on $T^7/Z_2$, which arises from D2-branes. T-dualizing three of these compact dimensions yields $T^3$-wrapped D5-branes on $T^3\times (T^4/Z_2)$. Due to the strong coupling present at points on $T^7/Z_2$, this argument is not entirely trustworthy, since we cannot trust our perturbative understanding of T-duality \cite{everybodyLies}. However, our ability to move the D-branes away from the strongly coupled $Z_2$ fixed points gives us some confidence that it is correct. More importantly, extra D-branes seem to be the only natural explanation for gauge symmetry enhancement.} all of the D5-branes are on equal footing, regardless of which compact dimension is responsible for their existence. Therefore, the 2d theory on our probe D1-brane is the same type of ADE theory as before, except it has an enhanced global symmetry. 

At first glance, nothing should change in this story if we compactify an additional circle. However, if we do so then the D1-brane and the D5-branes reside in the same number of non-compact (or rather, since we have compactified the spatial dimension of the D1-brane on a large circle, ``large") dimensions. Thus, the symmetry on the D5-branes is gauged and the Coulomb branch describes the motion of interacting effective D-strings, rather than the motion of a single D1-brane probe of a fixed (non-backreacting) background. This need to account for the extra gauge symmetry is familiar from studies of D1/D5 bound states describing $SU(N)$ instantons on K3\cite{vafa:k3inst,vafa:k3insts} or $T^4$\cite{sen:intersecting,vafa:hagedorn,w:higgs,dvv:t4inst2,dvv:t4inst,mms}, which focus on the Higgs branches of the relevant gauge theories.

There is another problem, hinted at in \cite{s:hlst}, which we have ignored in compactifying the dimensions transverse to the D1-brane: when there are one or two transverse spatial directions, the D1-brane we are studying suffers from the infrared sickness, common to all charged particles in one or two spatial dimensions, that results from asymptotic freedom \cite{sen:strong}. We can resolve this issue by compactifying to two dimensions, where there are no non-compact directions transverse to the D1-brane in which fields may diverge, but this only creates a new problem: the D1-brane filling the large dimensions creates a tadpole, analogous to the D9-brane tadpole of type I in 10 dimensions. In the type IIA or heterotic frames, we have a similar problem, where the tadpole is now associated with the B-field instead of R-R fields. The couplings studied in \cite{Lerche:1987qk}, in the heterotic case, and \cite{Vafa:1995fj,Sethi:1996es}, in the type IIA case, do not help us, as their contributions to the tadpole vanish. For example, this follows in the type IIA frame from the vanishing of the elliptic genus of K3$\times T^4$.

Fortunately, returning to the heterotic frame and taking $g_s\to0$ resolves both of these problems. In order to explain this claim, we first review some salient aspects of little string theory, and in particular the decoupling limit $g_s\to 0$ by which it is defined. As in \cite{s:hlst}, we begin by determining the coupling constant on the heterotic 5-brane by employing S-duality. Using the relations $g_s\to 1/g_s$ and \eqref{eq:weylTrans} that relate the type I and heterotic frames, the D5-brane worldvolume coupling constant $g_{5d}^2=\frac{g_s}{M_s^2}$ maps to the heterotic 5-brane worldvolume coupling $g_{5}^2=\frac{1}{M_s^2}$. Since the latter coupling is independent of $g_s$, we find that the worldvolume hosts an interacting theory even when $g_s\to 0$. However, \cite{s:hlst} emphasized that we could only trust this reasoning because string theory with a 5-brane background makes sense. In our case, we need to be a bit more careful, as the anomaly ruins this reasoning. We deal with these infrared issues, as in \cite{PPV}, by noting that the limit $g_s\to 0$ eliminates the backreaction of the 5-brane on its background. The tadpole tells us that we are unable to couple our 5-brane to a dynamical B-field, but we are still able to consider the 5-brane worldvolume theory in a fixed background. (This is analogous to the fact that we can study 2d or 6d field theories with a gravitational anomaly so long as we decouple them from gravity.) That is, when $g_s$ is small but finite there is a separation of scales between $M_s$ and the Planck scale, $M_p$, and the theory on the 5-brane is perfectly well-defined as an effective theory with a cutoff of $M_p$. Since the decoupling limit takes $M_p$ to infinity, our HLST is a UV-complete theory.

This allows us to address the issues we raised above. The low-energy gauge theory description of the compactified HLST is the same as the D1 worldvolume theory described above, except now nothing disastrous happens when we compactify down to two dimensions. In particular, by taking $g_s\to0$ we obtain a separation of scales between the $SU(2)$ coupling constant on the 5-brane and the spacetime coupling constant, which allows us to meaningfully distinguish between the gauge and global symmetries on the 5-brane. Furthermore, this HLST decoupling limit allows us to avoid problems from IR divergences when there are one or two non-compact directions transverse to the 5-brane. The way in which this happens is rather interesting. Although it is the case that for any non-zero value of the coupling constant the 5-brane adds infinite energy to the vacuum, it is also the case that for any fixed point in spacetime we can choose the coupling constant to be small enough that the effects of the 5-brane are negligible at this point. In this sense, the decoupling limit is pushing the IR divergence off to infinity. In field theory language, we are simply tuning a relevant coupling constant to zero. Finally, when we compactify all of the directions transverse to the HLST, the $g_s\to0$ limit allows us to deal with the tadpole, similarly to \cite{PPV}. (However, unlike \cite{PPV} we still have an interacting theory.) In the same way that this limit eliminates backreaction on the gauge fields, it also eliminates backreaction on the B-field.

\bigskip

Next, as described in section \ref{sec:origin}, we wish to situate ourselves at a point in moduli space that is slightly deformed away from a Niemeier point, in order to obtain an HLST with an Umbral global symmetry group. This deformation is actually necessary in order for the preceding discussion of the degrees of freedom near the singularities to be valid -- precisely at a point with enhanced global symmetry, the IR limit where we restrict to the Coulomb branch becomes singular \cite{aspinwall:theta,w:comments}. This is analogous to the fact that the type IIA K3 NLSM is only well-defined away from the singular points.\footnote{However, observe that no such singularity arises in the type IIA $T^4$ NLSM when the $T^4$ factor of K3$\times T^4$ yields enhanced gauge symmetry. This seems to be a counterexample to the principle articulated in \cite{w:various,aspinwall:theta,w:comments,w:evidence,w:smallInst} that perturbation theory breaks down when it is unable to correctly yield the massless spectrum. In particular, the mechanism studied in \cite{w:evidence} -- where perturbation theory for type I on $S^1$ at a point with enhanced gauge symmetry breaks down as a result of infrared divergences in the integration over worldsheet moduli -- does not seem to be available to us, as we have neither tadpoles nor open string worldsheets. Instead, it appears that perturbation theory fails rather silently, in that one could blissfully employ it to obtain wrong answers. The results of \cite{Antoniadis:1997zt,pioline:triality,pioline:3d} suggest that spacetime instantons that wrap a cycle of K3 times a cycle of $T^4$ make significant corrections to perturbation theory. In particular, the threshold corrections studied in these papers diverge at points of enhanced gauge symmetry \cite{pioline:diverge,Angelantonj:2012gw}. It would be interesting to use these results to better understand the breakdown of type II perturbation theory. Finally, we note that this appears to be a realization of the principle articulated in \cite{everybodyLies}: non-perturbative effects can make the physics of string theory on a torus depend on the background transverse to the torus. In fact, the connection with \cite{everybodyLies} is even stronger: if we start on $T^4\times T^4/Z_2$ at a point where enhanced gauge symmetry comes from the $T^4$ and we T-dualize one of the circles of the second $T^4$, we would na\"ively end up in IIB on $T^4\times T^4/Z_2$ where the $Z_2$ is generated by the product of $(-1)^{F_L}$ and reversal of all coordinates on the $T^4$. We know from above that enhanced gauge symmetry in this duality frame arises from the presence of extra NS5-branes (beyond the NS5-brane charge localized at the $Z_2$ fixed points), but if we trust T-duality then no such NS5-brane appears. We thus learn that the action of T-duality is non-perturbatively modified.} The upshot of this discussion is that we should avoid sitting precisely at points of enhanced gauge symmetry. On the heterotic side, this lifts the Higgs branch (except for the center of mass moduli that we have hitherto ignored), so that we are forced onto the Coulomb branch. Specifically, we sit at a point in moduli space where the masses of the hypermultiplets and the spacetime gauge bosons (the duals of the D1-D5 and D5-D5 fundamental strings, respectively) are of order $M_s\epsilon$.

The $g_s\to 0$ limit generically makes the little string theory insensitive to the geometry of the transverse dimensions, as the part of the moduli space describing motion in these directions decompactifies \cite{s:ns5}.\footnote{\label{ft:noncompact}While the discussion in \cite{s:ns5} was specific to the type II case, the conclusions hold equally well in the heterotic case. We can derive them as in \cite{s:ns5}, by considering the supergravity origin of the 5-brane hypermultiplet zero-modes \cite{strominger:hetSol,Harvey:instanton}. However, in keeping with the spirit of this section we first focus on the worldvolume description of the zero-modes. As discussed in Appendix \ref{sec:fieldReview}, the natural normalizations of the vector multiplet and hypermultiplet zero-modes of a D$p$-brane differ by a factor of $M_s^4$; in particular, this is true for both type I and type IIB compactified D$5$-branes. Strong-weak duality yields heterotic or type IIB 5-branes. This transformation modifies the vector multiplet coefficients via $\frac{M_s^2}{\tilde g_s}\to M_s^2$ and the hypermultiplet coefficients via $\frac{M_s^6}{\tilde g_s}\to \frac{M_s^6}{g_s^2}$. We thus reproduce the factor of $1/g_s^2$ that was crucial in \cite{s:ns5}; as $g_s\to 0$, this factor forces the hypermultiplets to only describe the immediate vicinity of the 5-brane. We note that this factor in the zero-mode action is consistent with the fact that the supergravity action of an instanton in the four dimensions transverse to a heterotic 5-brane is proportional to $1/g_s^2$.} 
The reason that the transverse $T^4$ is able to affect our HLST is clearly that the $T^4$, although classically smooth, has a size of order the string scale. Thus, this mechanism for generating little string theories is similar to that of \cite{intriligator:singular4}, where a 5-brane is situated near a geometric singularity; however, the theories we obtain are distinct. In particular, our low-energy field theories have no tensor multiplets.

Since at points of enhanced gauge symmetry the $T^8$ generally does not factor as $T^4\times T^4$, one should no longer expect the moduli space to be a product of center of mass and Coulomb factors. Furthermore, the moduli space is compact, as we explain in the next section. However, near the (resolved) singularities of the moduli space, we still locally have a factorization of the form $\RR^4\times$Coulomb if we take $\epsilon\ll 1$. One way to see this is that in this region, the 2d gauge theory description of the moduli space is accurate -- the light gauge theory degrees of freedom that we have integrated out are far more important than corrections to the gauge theory's moduli space coming from KK and stringy modes. This is the common phenomenon -- discussed in more detail in Appendix \ref{sec:fieldReview} -- where the moduli space decompactifies when we restrict to the field theory.

After taking the $g_s\to 0$ limit to define our HLST, we still want to take the low energy limit in order to obtain a 2d CFT. In particular, it is important to restrict to energies below $M_s\epsilon$ so that the light hypermultiplets we have integrated out do not give an uncontrolled non-local theory in the regime where we would like to trust the Coulomb branch sigma model. In the IR limit we thus obtain a non-singular CFT. We note that our ability to resolve the singularities of the Coulomb branch with hypermultiplet mass parameters is strongly dimension-dependent: in 4d or 5d, the presence of singular points in the moduli space is unavoidable. This can be traced back to the fact that the 2d gauge theory is related to type IIA on K3 and the 3d gauge theory is probing M-theory on K3, where the singularities can obviously be resolved by moving around in moduli space, whereas the 4d theory is probing F-theory on K3, where the elliptic fibration will always have singular fibers even if the total space is smooth.

\bigskip

We have finally completed the construction of our HLSTs and CFTs with Umbral symmetry. We now relate them back to Umbral moonshine. To do so, we note that the modified elliptic genus of our CFTs is always equal to $\hat Z^{T^4}Z^{K3}$. This follows either from duality with the IIA fundamental string or from computing as in section \ref{sec:generic} at a point where the moduli space simply factors as $T^4/Z_2\times \RR^4$. (In order to compute the modified elliptic genus, we regard the latter factor as the decompactification limit of a $T^4$.) Furthermore, near a point of enhanced gauge symmetry the modified elliptic genus obtains contributions of the form $\hat Z^{T^4}Z^{X,S}$ from the (resolved) singularities, since as we mentioned above the moduli space factors as $\RR^4\times$Coulomb near the singularities. We thus begin to see how the Umbral moonshine decompositions of the K3 elliptic genus will be explained. In order to make this explanation clearer, we now transition to a new duality frame.

\subsection{Static Gauge Description of a Flux Compactification}

As we mentioned above, there is a duality frame where the moduli space of the HLST -- and in particular, the collection of NS5-branes -- is the background seen by a fundamental string \cite{vafa:dManifolds,kutasov:orbSolitons,Sen:1996na}. A perturbative description of this duality frame is as an asymmetric orbifold of IIB on $T^4\times T^4$ by the $Z_2$ symmetry generated by the product of $(-1)^{F_L}$ and reversal of all coordinates on the second $T^4$. This is T-dual to IIA on $T^4\times K3$ and S-dual to a $T^4\times T^4/Z_2$ type IIB orientifold.\footnote{That mapping from $SO(32)$ heterotic to type I and then acting with S-duality should leave the coupling constant invariant led to the related proposal that $SO(32)$ heterotic string theory can be obtained as a perturbative orbifold of type IIB \cite{hull:ns9,hull:orbToHet,hull:orbToHet2}. This is complicated by the transformation properties of 9-branes under S-duality \cite{roo:IIB9}. It is argued in \cite{Sen:1996na} that compact dimensions play an important role in this argument, and that without the more complicated $Z_2$ action involving them S-duality need not hold for the orbifold.} Of course, we are really interested in non-perturbative points in moduli space where we have enhanced gauge symmetry coming from extra NS5-branes and where the geometry does not simply factorize. (We say extra because even at generic points in moduli space we have 16 NS5-branes.) These points have B-field and dilaton backgrounds turned on, and perturbation theory is unhelpful in describing them. Nevertheless, because we have no R-R backgrounds turned on we can formulate the sigma model of a static gauge string. In fact, this is precisely the low-energy limit of the HLST that we studied in the previous section. The NS5-branes appear in the sigma model as points in the target space about which the B-field background undergoes monodromies. We note that similar flux compactifications were studied in \cite{mcGreevy:nonGeo} (see also \cite{vafa:dManifolds,ooguri:ADE}), and many of the above observations were made in this context. In this duality frame, the anomaly is manifestly unproblematic, as we are simply studying the perturbative quantization of a string in a non-anomalous background. However, we remind the reader that the Coleman-Mermin-Wagner theorem makes the notion of choosing an Umbral point in moduli space problematic. One can therefore view the Umbral symmetries as perturbative symmetries or, since we are not actually interested in string interactions, as symmetries of the theory on the string worldsheet.

Note that it is important that the usual $\RR^4$ factor of the HLST moduli space becomes compactified and merges with the Coulomb branch, since the background obtained by pairing the geometry at a generic point in moduli space with enough NS5-branes to yield a rank 24 non-Abelian gauge symmetry is anomalous.

The NS5-branes in this duality frame allow us to clarify a number of claims we made in the previous section. Specifically, we now explain why the modified elliptic genus splits as a sum of two terms, one of which is $\hat Z^{T^4}Z^{X,S}$, the contribution from an NS5-brane throat times $\RR^4$. The point is simply that because NS5-branes are so singular, they are insensitive to the background that surrounds them. Indeed, when one studies a string background that contains NS5-branes, one finds that the entire throat is squished to a point \cite{kutasov:orbSolitons}!\footnote{A related squishing of NS5-branes is discussed in \cite{w:comments}.} The throat is then thought of as the holographic description of the NS5-brane \cite{s:ns5holo}, in the sense of open-closed duality that is familiar for D-branes. It essentially pushes the core of the NS5-brane far away from the rest of spacetime \cite{w:comments}. So, as we proceed down the throat the effects from the outside of the throat become arbitrarily negligible \cite{maldacena:throat}. Of course, we have resolved the singularity by Higgsing to an Umbral point in moduli space, so this infinite throat description is slightly modified. But, the limit described in the previous section, with $\epsilon$ fixed, guarantees that the NS5-branes are still close enough to notice each other.

The Umbral moonshine decompositions of the K3 elliptic genus are now explained by the following:
\be\label{eq:division} \hat Z^{T^4\times K3}=\hat Z^{T^4}Z^{K3}=\hat Z^{T^4}Z^{X,S}+\hat Z^{\text{rest}} \Rightarrow \hat Z^{\text{rest}} = \hat Z^{T^4} Z^{X,U} \ . \ee
Geometrically, this decomposition corresponds to gluing the non-compact throat together with the non-compact rest of the background in order to form the compact target space. This non-compactness explains the role of mock Jacobi forms in Umbral moonshine. We can also see why $Z^{X,S}$ includes only the discrete part of the spectrum: the continuous part of the spectrum is eliminated by our deformation to the Umbral point and gluing together with the rest of spacetime. This gluing is related to the `semi-flat approximation' used to construct non-singular string backgrounds in \cite{mcGreevy:nonGeo,strominger:mirrorT}. It is also related to the combination of twisted and untwisted sectors in \cite{kutasov:orbSolitons}.

Similar reasoning allows us to relate K3 twining genera to indices in our CFT, as described in \S\ref{sec:twining}. The symmetries act on the throat CFTs by permuting the simple roots of the spacetime gauge symmetry, whose corresponding gauge bosons are furnished by D1-branes strung between the NS5-branes \cite{vafa:dManifolds}. This is rather interesting from the 2d gauge theory point of view, since the vector multiplets whose scalars parametrize the Coulomb branch are uncharged under these symmetries, but the Umbral groups nevertheless act non-trivially on the throat CFTs.\footnote{See \cite{sw:3d} for a related discussion in 3d.} We thus see that the hypermultiplets leave their mark on the Coulomb branch in the form of discrete bound states localized near the origin \cite{dvv:2dbh,ooguri:ADE,s:coulomb,w:higgs,brodie:2dMirror,aharony:LSTmatrix,harvey:quantumHiggs}. In particular, these worldsheet states yield spacetime scalars (that is, under the 6d Lorentz group) that correspond (i.e., holographically) to the gauge-invariant polynomials in the scalars of the 6d $\N=(1,1)$ vector multiplet that describes the NS5-branes at low energies \cite{s:ns5holo}.

\bigskip

Actually, while this gluing picture is appealing, it is not clear that the Umbral part of the target space can be separated from the NS5-brane throats in order to define an independent non-compact Umbral NLSM. What we can say with certainty is that near the points with enhanced gauge symmetry, where the inside of the throat decouples from the rest of spacetime, discrete states localized inside the throats do not mix with the rest of the spectrum under the action of the Umbral groups. In this more conservative approach, the mock automorphic forms arise because we are only considering the discrete states in the throat CFTs, since after all the entire target space is compact.

However, there is reason to hope that the Umbral NLSMs exist. One can imagine stretching the throat to be arbitrarily long, by approaching the point with enhanced gauge symmetry, so that its tip is extremely distant from the Umbral part of the target space; this picture seems to suggest that an independent Umbral CFT will emerge in this limit. One might worry that such a limiting CFT would be singular, as is the case for the throat CFT \cite{w:comments}. However, the intuition for what goes wrong in the latter case does not apply to the former. While the target space becomes non-compact, this process is smooth -- a point is not discontinuously replaced by an infinite throat. Furthermore, the dilaton need diverge only deep inside the throat, not at the outside of the throat where the Umbral target space resides. Therefore, it is conceivable that the worldsheet theory at a point with enhanced gauge symmetry can be divided into a singular throat CFT and a non-singular Umbral CFT (whose modified elliptic genus is $\hat Z^{\text{rest}}$).\footnote{However, it is known that the opposite process of joining two throats to form an NLSM can fail \cite{w:comments,sk:decoupling,w:higgs,aspinwall:theta,aharony:LSTmatrix,verlinde:higgsCoulomb,w:holoHiggs}.} The Umbral groups are still the relevant symmetry groups at these points in moduli space; in contrast with \S\ref{sec:origin}, the quotient by the Weyl group now arises (as in \cite{slowpaper}) because the latter consists of gauge symmetries.

\section{Toward Antumbral Moonshine}\label{sec:ant}

Nothing in our reasoning depended on the non-Abelian gauge symmetry having rank 24. By considering points in moduli space with gauge symmetry characterized by an ADE root lattice $X$ whose rank is less than 24, we therefore discover that it is natural to expand the K3 elliptic genus as $Z^{X,S}+Z^{X,A}$, where $Z^{X,S}$ is the contribution from the ADE singularity specified by $X$ and $Z^{X,A}$ is termed the Antumbral contribution. In particular, we can consider points in moduli space where up to four of the dimensions decompactify. The question is then if $Z^{X,A}$ has a nice number theoretic construction, in analogy with that of $Z^{X,U}$ in the Umbral cases. We leave it to future work to settle this question, although we speculate on the connection between the physics in this paper and the number theoretic properties of $Z^{X,U}$ (and possibly $Z^{X,A}$) in the conclusion. Here, we content ourselves with determining some of the Antumbral symmetry groups $G^X$ and the representations we obtain by decomposing those of Umbral moonshine. Of course, we no longer have the decomposition $\Gamma^{24,8}=L^X\oplus E_8(-1)$, so we should be more precise about the definition of the Antumbral groups. We retain the definition $G^X=O(L^X)/\Weyl(X)$ by defining $L^X=\Gamma^{24,8}\cap (\RR\otimes X)$. However, we note that $L^X$ and $G^X$ now depend not only on $X$, but also on its embedding into $\Gamma^{24,8}$ (so we are abusing notation a bit). The reasoning of section \ref{sec:origin} demonstrates that $G^X$ is the maximal group that acts non-trivially on $X$ and is preserved by an infinitessimal motion in moduli space that breaks the non-Abelian gauge group to its Cartan torus. Note that $L^X$ will, in general, be even and positive definite but not unimodular. As in \cite{Cheng:2013wca}, we learn that $G^X$ has a natural permutation representation that acts on a set of simple roots for $X$.

For simplicity, we focus on cases of the form $X=A_1^n$, where $Z^{X,S}=n Z^{A_1,S}$. The Antumbral group is then always a subgroup of $S_n$ (the symmetric group), but it is in general not equal to $S_n$ because of the presence of glue vectors which lie outside the root lattice (see e.g. \cite{ConwaySloane}). However, when $n\le 8$ we can choose points in moduli space such that $\Gamma^{24,8}$ decomposes as $(\Gamma^{1,1})^n\oplus \Gamma^{24-n,8-n}$. Since $\Gamma^{1,1}$ is spanned by two orthogonal vectors, $u_1$ and $u_2$, such that $u_1^2=-u_2^2=2$, we can write $\Gamma^{1,1}\cong A_1\oplus A_1(-1)$. So, when $n\le 8$ we can decompose $\Gamma^{24,8}$ as $A_1^n\oplus (A_1(-1))^n\oplus \Gamma^{24-n,8-n}$, which demonstrates that we can choose $G^{A_1^n}=S_n$. At the other extreme, we now consider $A_1^{23}$. We proceed as in section \ref{sec:origin}, by beginning at the $A_1^{24}$ Niemeier point and deforming away. Specifically, we pick one of the 24 roots, $r$, and find the subgroup of $O(L^{A_1^{24}})$ (thought of as a subgroup of $O(\RR \otimes L^{A_1^{24}})$) that fixes both $r$ and the Weyl vector of the $A_1^{23}$ root lattice that is orthogonal to $r$. (Note that nothing forces $\epsilon$ to be infinitessimal in \eqref{eq:changebasis}, so we can move far away from the $A_1^{24}$ point if we wish.) Working with the MAGMA programming language, we find the sporadic Mathieu group $M_{23}$. Note that this embeds into $G^{A_1^{24}}=M_{24}$, since the elements of this group fix the $A_1^{24}$ Weyl vector. When we restrict $M_{24}$ to $M_{23}$, we have $\bf{45}\to \bf{45}$, $\bf{231}\to\bf{231}$, and $\bf{770}\to\bf{770}$. Similarly, we find points with $X=A_1^{22}$. To do so, we again begin at the $A_1^{24}$ Niemeier point, pick two orthogonal roots, $r_1$ and $r_2$, and find the (pointwise) stabilizer of the set consisting of $r_1,r_2$, and the Weyl vector of the $A_1^{22}$ root lattice orthogonal to both $r_1$ and $r_2$. This yields $G^X=M_{22}$. If we instead only require that $r_1+r_2$ be stabilized, but not $r_1$ and $r_2$ individually, then we find $G^X=M_{22}$:2. When we restrict $M_{24}$ to $M_{22}$:2, we have $\bf{45}\to \bf{45}$, $\bf{231}\to\bf{231}$, and $\bf{770}\to \bf{210}+\bf{560}$.

\section{Eternal Moonshine of the String Theorist Mind}\label{sec:conclude}
\epigraph{$J_3$ seems unrelated to any other sporadic groups (or to anything else).}{\textit{Wikipedia contributor R.e.b.}\cite{janko3}}

In this paper, we have set forth a physical explanation for a number of results in \cite{Cheng:2014zpa}. We conclude by suggesting some directions for further research.

\begin{itemize}
\item While we have provided the physical setup in which Umbral moonshine seems to be naturally realized, this does not mean that we have a particularly useful description of the non-compact Umbral sigma models. In fact, it is not even certain that they exist on their own -- they may only make sense when glued to the NS5-brane throats. We speculated on a means for constructing these theories, but it would be desirable to know if they actually exist. The existence of the Umbral moonshine modules may support this hypothesis. On a related note, it is not immediately clear how to extract the Umbral mock modular forms $H^X$ from our current picture, although they are clearly present. The simple relation between $H^X$ and $Z^{X,U}$ suggests that these CFTs, if they exist, may be formulated more directly. One hint, noted in \cite{Cheng:2014zpa}, is provided by the fact that the operation in \eqref{eq:zUmbral} resembles the orbifolding operation of \cite{vafa:orbifold,kawai:orbifold} by $e^{2\pi i J_0}$. Another hope is that the explicit construction of some Umbral moonshine modules \cite{Harvey:E8,duncan:someAs,cheng:someDs,cheng:d46} will inform the study of the Umbral sigma models. (This may, in turn, lead to a general construction of all Umbral moonshine modules.) Finally, the relationship between these theories and the IR limits of compactified 5-branes may prove to be quite useful. For example, a DLCQ approach along the lines of \cite{kachru:dlcq,lowe:dlcq} may prove fruitful.
\item It would be interesting to find a physical motivation for the construction in \cite{Cheng:2013wca} of the mock modular forms $H^X_g$. From a number theoretic point of view, these functions are very special\cite{Cheng:2011ay,cheng:optimal,Cheng:rademacher}. In particular, they have an intimate relationship with Rademacher sums. Such a relationship had previously arisen in a holographic context\cite{moore:farey}, so it was suggested in \cite{Cheng:2011ay} that Umbral moonshine may have an intimate relationship with holography. Our invocation of heterotic little string theory may provide a hint that this is correct, since these theories have holographic duals \cite{kapustin:hlst,narain:hlst}; these dualities are unique in that there are a number of quantities we may compute on both sides of the duality, even for a single 5-brane! In particular, these references demonstrated a holographic computation of part of the spectrum of BPS observables; applying these techniques to the compactified case with enhanced gauge symmetry discussed in this paper could yield interesting results for moonshine. Indeed, computations along these lines were recently performed for type II 5-branes, and the state counts that arose were related to the mock modular forms of Umbral moonshine \cite{Harvey:2013mda,Harvey:2014cva}. It was noted in \cite{s:ns5holo,kapustin:hlst} that because of the strongly-coupled region in the bulk, this perturbative analysis misses part of the spectrum. One might hope that the missing states in this case would be those counted by $\hat Z^{T^4} Z^{X,S}$, and that one would be left with Umbral functions; the results of \cite{Harvey:2013mda,Harvey:2014cva} could then be taken as reason for optimism.
\item Given any Umbral symmetry that fixes a positive-definite 2-plane, we can decompactify to four dimensions and retain this symmetry. It would be interesting to try to build CHL models involving orbifolds by these symmetries -- in particular, the 1/4-BPS dyons in such theories would likely be counted by multiplicative lifts of the Umbral twining genera, as described in \cite{mz:walls}. Similarly, 1/4-BPS dyons in CHL-like models associated to positive-definite 2-plane preserving symmetries that are not in any Umbral group would likely be counted by (some of) the functions proposed in \cite{mz:walls} that were conjectured to be related to an extended Conway moonshine.
\item We have motivated the study of Antumbral decompositions of the K3 elliptic genus. It would be appealing to understand if the functions $Z^{X,A}$ have special number theoretic characterizations. For example, one might hope that the 16 weight one optimal mock Jacobi forms with rational coefficients that do not play a role in Umbral moonshine \cite{cheng:optimal} do play a role in an Antumbral picture. (On the other hand, this would contradict the intuition we suggested above involving Rademacher sums.) If these functions are uninteresting, one would hope that a physical explanation for the number theoretic aspects of Umbral moonshine would be able to explain why this is the case. Of course, this is clear from the representation theory side of moonshine -- in comparison with the Umbral case, many more states counted by $Z^{X,A}$ do not transform under the Antumbral groups. 
\item The functions $Z^{X,S}$ are related to the polar part of a meromorphic Jacobi form \cite{Cheng:2014zpa}. Such functions exhibit the phenomenon of wall crossing \cite{Dabholkar:2012nd}; this may play an important role in a complete understanding of Umbral moonshine.
\item Second quantization frequently leads to lifts of automorphic forms which have interesting physical interpretations (see, e.g., \cite{DVV,harrison2016heterotic,mz:walls,PPV}). Such a procedure may be straightforward to implement here, as well, since generalizing from a single heterotic 5-brane to $k$ such branes simply changes the worldvolume gauge symmetry from $Sp(1)$ to $Sp(k)$. This could lead to an interesting notion of a mock Siegel form,\footnote{We thank Arnav Tripathy for suggesting this possibility.} and may also suggest relationships between Umbral moonshine and generalized Kac-Moody algebras.
\item Compactified little string theories may provide a useful setting for studying sigma models whose associated string theories have infrared problems. For example, just as heterotic-type IIA duality relates the fundamental IIA string to a compactified heterotic 5-brane, it also relates the fundamental heterotic string to a compactified NS5-brane. This may enable an interesting reformulation of the Monstrous moonshine results of \cite{PPV}, since the free second-quantized heterotic strings studied therein may have a dual description in terms of an interacting little string theory -- the zero coupling limit imposed by anomalies does not preclude the existence of an interesting interacting limit! The above speculations on Rademacher sums and 5-brane holography are equally applicable in this case \cite{s:ns5holo}, since the genus zero property of Monstrous moonshine may be reformulated using Rademacher sums \cite{MR2905139}. Similarly, the authors of \cite{harrison2016heterotic} were hesitant to interpret their second quantized sigma model in its associated string theory context, as the meaning of counting charged states in low dimensions was unclear. However, if we reinterpret their fundamental heterotic strings as type IIA NS5-branes then the gauge symmetries of spacetime become global symmetries of the little string theory, and there is no problem with states in low dimensions that are charged under a global symmetry!
\item We now have a rich web of relations between automorphic forms, representation theory, and string theory \cite{Conway:1979kx,FLM,PPV,Eguchi:2010ej,Cheng:2012tq,Cheng:2013wca,Cheng:2014zpa,harrison2016heterotic,duncan:2014eha,duncan:2015xoa}. However, it is certain that the nature of these relationships -- especially those relating different moonshines (and, in particular, Conway and Umbral moonshines) -- has yet to be fully explored. (Although, see \cite{mz:walls,slowpaper,duncan:2015xoa,duncan:selfDual,wendland:conwayK3} for recent progress in relating Conway and Umbral moonshines.) Since it has proved natural to study many of these from the perspective of low-dimensional heterotic string compactifications, it seems reasonable to hope that these compactifications will provide an organizing principle, the study of which could yield important insights for both mathematics and physics. For example, the prospect that string theory could provide an intuitive reason for why the sporadic groups exist is exciting since (as the quote at the start of this section indicates) such a justification is, at the moment, sorely lacking. Conversely, the appearance of these groups as symmetries of some of the simplest string compactifications may hint at some hidden structure in string theory.
\end{itemize}

\section*{Acknowledgments}
\epigraph{I like pigs. Cats look down on you; dogs look up to you; but pigs treat you like an equal.}{\textit{Winston Churchill}}
We extend our profound gratitude to S. Kachru for interesting and helpful conversations, and for his enduring support and positivity. We additionally thank A. Adams, S. Harrison, N. Paquette, B. Pioline, A. Tripathy, and R. Volpato for enjoyable conversations. Finally, we thank S. Harrison, S. Kachru, and N. Paquette for comments on an earlier draft of this paper.

\appendix

\section{Special Functions}\label{a:basics}

Readers are referred to \cite{Cheng:2013wca} for mathematical background. We define $q= e^{2 \pi i \tau},$ $y= e^{2 \pi i z}$.

The Dedekind eta function of weight $\half$ is
\begin{equation}
\eta(\tau) = q^{\frac{1}{24}} \prod_{n=1}^{\infty} (1-q^n)=q^\frac{1}{24}\sum_{n=-\infty}^\infty (-1)^n q^\frac{3n^2-n}{2}=q^{1/24}(1-q-q^2+O(q^3)).
\end{equation}

The classical theta functions are Jacobi forms of weight $\half$ and index $1$ and can be written as follows:
\begin{align}
\theta_1(\tau,z) 
&= -i q^{\frac{1}{8}} y^{\half} \prod_{n=1}^{\infty} (1-q^n)(1-y q^n)(1 - y^{-1} q^{n-1}) \\  
&= i \sum_{n=-\infty}^{\infty} (-1)^n q^{\frac{(n-\half)^2}{2}} y^{n-\half} \nonumber \\
\theta_2(\tau,z) 
&=q^{\frac{1}{8}} y^{\half} \prod_{n=1}^{\infty} (1-q^n) (1+y q^n) (1 + y^{-1} q^{n-1}) \\
&= \sum_{n=-\infty}^{\infty} q^{\frac{(n-\half)^2}{2}} y^{n-\half} \nonumber \\
\theta_3(\tau,z) 
&= \prod_{n=1}^{\infty} (1-q^n) (1 + y q^{n-\half}) (1 + y^{-1} q^{n-\half})  \\
&= \sum_{n=-\infty}^{\infty} q^{\frac{n^2}{2}} y^n \nonumber \\
\theta_4(\tau,z)
&= \prod_{n=1}^{\infty} (1-q^n) (1 - y q^{n-\half}) (1 - y^{-1} q^{n-\half}) \\
&= \sum_{n=-\infty}^{\infty} (-1)^n q^{\frac{n^2}{2}} y^n \nonumber.
\end{align}

We also write the standard Jacobi forms $\chi_{0, 1}(\tau, z)$ of weight $0$ and index $1$ and $\chi_{-2, 1}(\tau, z)$ of weight $-2$ and index $1$ \cite{eichler_zagier}:
\begin{align}
\chi_{0, 1}(\tau, z)&= 4\left(\sum_{i=2}^4 \frac{\theta_i(\tau, z)^2}{\theta_i(\tau, 0)^2} \right) = (y^{-1}+10+y)+O(q) \\
\chi_{-2, 1}(\tau, z)&= -\frac{\theta_1(\tau, z)^2}{\eta(\tau)^6} = (y^{-1}-2+y)+O(q).
\end{align}

\section{Field Theory Probes of Stringy Phenomena}\label{sec:fieldReview}

In this appendix, we elaborate on a number of statements in \S\ref{sec:duality}.

\subsection{Generic Points in Moduli Space}\label{sec:generic}

We first review the discussion in \cite{w:smallInst} that explains how to map the BPS degrees of freedom of the fundamental type IIA string on K3 to those of its heterotic dual on $T^4$. We restrict ourselves, in this section, to generic points in the moduli space of the compactified $SO(32)$ heterotic string, where the string compactification's gauge group is broken to its maximal Abelian subgroup, $U(1)^{24}$. The essential point is that this string is to be regarded, in the heterotic frame, as the `zero-thickness' limit of a 5-brane soliton wrapped around $T^4$.\footnote{Technically, the heterotic 5-brane is dual to a type IIA string in static gauge \cite{Harvey:soliton,sen:soliton}.} More precisely, the gauge fields (and their superpartners) in the four dimensions transverse to the 5-brane describe a small instanton of $\N=4$ Yang-Mills, where the gauge group of this field theory is the gauge group of the string theory. This singularity in the gauge bundle allows for a non-perturbative enhancement of the 5-brane worldvolume's gauge symmetry.\footnote{We emphasize the difference between this case, where our non-perturbative enhanced gauge symmetry comes from the singular gauge bundle, and that of type IIA, where enhanced gauge symmetry comes from singularities in spacetime. In fact, there are perturbatively controlled heterotic compactifications with ADE singularities in the spacetime background \cite{sen:hetAn,w:ade}.} Specifically, $k$ of these coincident fivebranes have $Sp(k)$ gauge symmetry in their worldvolume. (This is the same gauge group that we obtain from $k$ coincident type I D5-branes, as is expected from the fact that the strong-coupling limit of our heterotic fivebrane is the type I D5-brane.\footnote{The reader might be puzzled by our requirement that the gauge symmetries of the dual theories must agree, since gauge symmetries are not required to be invariant under dualities. However, in this case it is clear that on both sides of the duality we require a $(1,0)$ field theory whose Higgs branch is the appropriate ADHM moduli space, and this forces the gauge symmetries on both sides to be the same.}) So, when $k=1$ we have $Sp(1)=SU(2)$ gauge symmetry. In the limit of zero heterotic string coupling, the worldvolumes of these branes are described by interacting heterotic little string theories (HLSTs) \cite{s:hlst}. In particular, even a single heterotic 5-brane has a non-trivial decoupling limit; this is to be contrasted with the type II case, where at least two NS5 branes are required\cite{s:ns5holo,kapustin:hlst}.

The heterotic 5-brane wrapped around $T^4$ yields an effective 1-brane with two types of moduli. First, there are Coulomb branch scalars which describe the Wilson lines of the $SU(2)$ gauge group around the $T^4$. Since $SU(2)$ has rank 1, this yields four real scalar degrees of freedom. In addition, there are Higgs branch moduli which describe the instanton in the transverse directions. By wrapping the 1-brane around a large circle, we obtain a low energy effective field theory with supersymmetric boundary conditions (i.e. periodic boundary conditions for both bosons and fermions) in the spatial direction. In accordance with the Coleman-Mermin-Wagner theorem, there is no moduli space at the quantum level; instead, we obtain a $(4,4)$ NLSM whose target space is what would, in higher dimensions, be the moduli space. (We will still call this space the moduli space.)

We can easily argue that it is the Coulomb branch that describes the degrees of freedom coming from the type IIA string probing the K3 of its spacetime. Note that the string theory's Abelian gauge group has no non-singular instantons. Therefore, the only Higgs branch moduli parametrize the position of the 5-brane in the four transverse directions -- that is, these are the center of mass moduli of the small instanton. In the IIA picture, they parametrize the position of the string in these directions. Furthermore, an explicit description of the Coulomb branch demonstrates that it is a K3 orbifold. For, the four scalars parametrize the space $T^4/Z_2$, where the quotient comes from the Weyl group of $SU(2)$. This is the group of gauge symmetries that map the maximal torus of $SU(2)$ to itself (via conjugation), modulo the maximal torus itself (which acts trivially). If we denote the elements of the maximal torus by $\twoMatrix{e^{i\theta}}{0}{0}{e^{-i\theta}}$, then a representative for the generator of the Weyl group is $\twoMatrix{0}{-1}{1}{0}$, which implements $\theta\to-\theta$. Since the center of mass fields do not interact with the rest of the theory, we obtain a factorized sigma model with target space $\RR^4\times T^4/Z_2$; this factorization allows us to isolate the second factor, which is the one that interests us.

Of course, the K3 background of the type IIA compactification is not always a $T^4/Z_2$ orbifold. We will discuss in the next section how quantum corrections modify this moduli space; at the generic points we are currently considering, we obtain a smooth surface. However, this does not mean that the moduli space is the K3 surface of the IIA background; this is made exceptionally clear near points of enhanced gauge symmetry where, as we discuss in the next section, the moduli space can be quite different from a K3 surface. This apparent discrepancy is resolved by noting that the points in moduli space where the type IIA and heterotic perturbative descriptions are valid differ. Therefore, we can only match BPS states -- which we can follow from strong to weak coupling -- across this duality. In particular, we can match the elliptic genus of $T^4/Z_2$ against the elliptic genus of the type IIA background's K3; these are, of course, equal, since $T^4/Z_2$ is a K3 orbifold. (We emphasize that we have discussed two independent arguments for the result that the elliptic genus of the compactified HLST equals $Z^{K3}$: one based on heterotic-type IIA duality, and one based on consideration of the Coulomb branch moduli space. In fact, one can simply view the type IIA string as motivation for considering the heterotic 5-brane and for anticipating the role of $Z^{K3}$, as all of our results may be derived in the heterotic frame.) Because of the boundary conditions imposed above, we compute the elliptic genus in the Ramond-Ramond sector.

Before we proceed, we wish to emphasize two important aspects of our setup. First, the fact that Coulomb-Higgs factorization allows us to isolate the degrees of freedom describing the type IIA string on K3 from those describing its motion in the transverse directions is essential. For, fermion zero modes cause the elliptic genus of the transverse directions to vanish; this remains true even if we toroidally compactify some of them.\footnote{Actually, as we discuss in footnote \ref{ft:noncompact}, a rescaling involved in the definition of the little string theory only allows the Higgs branch to probe the immediate vicinity of the 5-brane, so even when dimensions transverse to the 5-brane are compactified in spacetime the corresponding dimensions in the sigma model target space remain non-compact. A similar effect does not occur for Coulomb branch moduli.} Second, the fact that the Coulomb branch is compact relied on the fact that we are considering the moduli space of the compactified HLST, as opposed to a 2d gauge theory on the effective 1-brane\cite{intriligator:compact}. For, without the infinite tower of momentum states in the compact directions the Wilson lines are not compact variables. This suggests that the 1-brane may be useful as a local probe of the HLST's moduli space (since locally the states in the tower are extremely heavy), but that it will miss the global features of the moduli space (since these generically-heavy states that we are ignoring can become light). We will see this explicitly in the next section.

\subsection{Points of Enhanced Gauge Symmetry}\label{sec:enhance}

We now investigate how the above story is modified at points in moduli space with enhanced gauge symmetry. The interesting complication that arises is that there is now a Higgs branch, and the Coulomb branch develops a singularity at its origin where the Higgs and Coulomb branches classically meet. Physically, this occurs because new states become light at this singularity, so the moduli space approximation breaks down. From the perspective of the 1-brane, $SU(2)$ Wilson lines are cancelling off the mass of some state that arises from its Wilson lines for the spacetime gauge group \cite{w:smallInst,porrati:H,sethi:commentH,s:K3}.

In order to study these singularities, we will find it more convenient to temporarily work in the strong coupling limit of the $SO(32)$ heterotic theory, where the appropriate perturbative description is type I on $T^4$. (This is not strictly necessary, but it will allow us to employ familiar techniques involving D-branes.) In this duality frame, the small instanton 5-brane is the D5-brane and the spacetime gauge symmetry arises from Chan-Paton factors on 32 D9-branes in the presence of an O9-plane. The heterotic 5-brane is dual to the type I D5-brane. This might seem surprising, given that the tension of an instanton 5-brane should scale with the Yang-Mills coupling as $1/g_{YM}^2$ and heterotic-type I duality inverts the string coupling, $g_s$ \cite{w:various}. The resolution of this involves two ingredients \cite{w:various,w:evidence}. First, in the heterotic theory we have $g_{YM}^2\sim g_s^2$, while in the type I theory we have $\tilde g_{YM}^2\sim \tilde g_s$ (we temporarily employ tilded variables in order to refer to type I parameters), since in the former the gauge bosons arise from closed strings while in the latter they arise from open strings. Second, in addition to $\tilde g_s= 1/g_s$ this duality also involves a Weyl transformation between the metrics of the two theories:
\be\label{eq:weylTrans} \tilde g_{\mu\nu} = g_s^{-1}g_{\mu\nu} \ .\ee
One way to arrive at this is by requiring the proper transformation of the Planck mass: $M_s^4/g_s\to M_s^4/\tilde g_s=M_s^4g_s$. Since the 10d Yang-Mills coupling has mass dimension $-3$, it follows that $g^2_{YM}\sim g_s^2/M_s^6$ must be rescaled by a factor of $1/g_s^3$ as we transform to type I variables, and we thus find that it scales in the same way as $\tilde g^2_{YM}\sim \tilde g_s/M_s^6 \sim 1/g_s M_s^6$. That is, these relations explain how a heterotic instanton 5-brane whose tension scales as $1/g_{YM}^2$ could be dual to a D5-brane whose tension scales as $1/\tilde g_{YM}^2$. To emphasize that this duality between instantons of dual theories is not automatic, we recall from the last section that the heterotic instanton 5-brane is dual to a fundamental type IIA string, and the tension of the latter certainly does not diverge as the type IIA string coupling vanishes. This is just another way of saying what we said before: the weakly coupled descriptions of dual objects do not have to match.

T-dualizing the directions of the $T^4$ turns Wilson lines into positions on the compact directions. After this duality transformation, we now have a type IIB orientifold on $T^4/Z_2$ with 16 O5-planes located at the fixed points of the $Z_2$ action. The heterotic 5-brane has become a D1-brane and the $SU(2)$ Wilson lines have become the position of the D1-brane in $T^4/Z_2$. Meanwhile, the D9-branes have become D5-branes whose positions are specified by what used to be the $SO(32)$ Wilson lines. This D1 probe of a D5/O5 configuration was studied in \cite{s:coulomb}; the analogous D2 probe of a D6/O6 configuration was studied in \cite{s:K3}; the D3 probe in \cite{s:fBranes,sw,sw:argyresDouglas,ganor:compact}; the D4 probe in \cite{s:5d,s:extremal,vafa:delPezzo}; and the D5 probe in \cite{w:smallInst,s:6d}. These theories were related, via compactification, in \cite{sw:3d,s:branesFields}. For a review of these results, see \cite{morrison:compactDual}. (For completeness, although these will not play a role in this paper, we mention that the D0 probe was studied in \cite{shenker:d0Short,diaconescu:d0d4} and the D-instanton / D3 system was studied in \cite{dorey:dInst}.) In all of these cases, the field theory on the D$p$-brane probe is super Yang-Mills with 8 supercharges and some number of hypermultiplets. Near an O-plane, the gauge group is $SU(2)$, and away from the O-plane it is $U(1)$. There is a neutral hypermultiplet that parametrizes the position of the probe along the directions in which the probed D$(p+4)$-branes extend; this decouples from the rest of the theory and will only play a small role in what follows. Finally, there are hypermultiplets associated to the D-branes being probed that transform in the fundamental representation of the gauge group. They transform under an $SU(N_f)$ global flavor symmetry when the gauge group is $U(1)$ and an $SO(2N_f)$ global symmetry when the gauge group is $SU(2)$; the $SO(2N_f)$ symmetry in the latter case reflects the presence of mirror D-branes on the opposite side of the O-plane. These global symmetries on the probes correspond to gauge symmetries of the string theory. The masses of these hypermultiplets correspond to the positions of the D$(p+4)$-branes in the directions transverse to their worldvolumes, while vector multiplet scalars correspond to the position of the probe D$p$-brane in these directions.\footnote{This statement, and some in the next paragraph, are modified slightly when $p=2$, as we describe below.} If we normalize our vector multiplets so that the kinetic terms for their bosonic components are, schematically,\footnote{We will be a bit careless with numerical factors. For a more careful treatment, see \S X.3.1 of \cite{instantonReview}.} $-\frac{1}{g_{p+1}^2}\Tr \brackets{\frac{1}{4}F_{\mu\nu}^2 + \frac{1}{2}(D_\mu \vec r)^2}$, then the vector multiplet scalars $\vec r$ are related to the physical position of the D$p$-brane via $\vec r=\vec x M_s^2$. A similar rescaling relates the physical positions of the D$(p+4)$-branes to the hypermultiplet mass parameters; in fact, the latter reside in background vector multiplets (where the corresponding gauge fields are those of the D$(p+4)$-branes). Finally, we normalize hypermultiplets so that their bosonic kinetic and mass terms are $-\frac{1}{g_{p+5}^2}\Tr\brackets{\frac{1}{2}(D_\mu H)^2 + \frac{1}{2}m^2 H^2}=-\frac{M_s^4}{g_{p+1}^2}\Tr\brackets{\frac{1}{2}(D_\mu H)^2 + \frac{1}{2}m^2 H^2}$, as is appropriate for describing the zero-modes of non-small instantons in the D$(p+4)$-brane worldvolume.

However, there is an important effect we must account for. As mentioned above, when we consider a $(p+1)$-dimensional gauge theory without a tower of Kaluza-Klein states, we should expect a non-compact moduli space. Geometrically, we can think of the size of the moduli space as being related to the KK scale, so that scaling out the KK modes blows up the size of the moduli space \cite{intriligator:compact}. Alternatively, we can note that it is only consistent to ignore the KK modes if we restrict to Coulomb branch vevs below the KK scale. This means that the probe gauge theories can only yield local information about the moduli space. In particular, the D1-brane gauge theory will be useful for studying degrees of freedom localized near the singularities of the Coulomb branch of the HLST, but in order to study global aspects of the Coulomb branch we will need to return to the sigma model of the previous section.

One important aspect of these $U(1)$ and $SU(2)$ gauge theories is that they can have an `anomaly.' The significance of these anomalies depends on the dimension of the probe \cite{s:6d,s:5d}. However, in all of these cases it has a simple spacetime interpretation: when the probe is a D$p$-brane, the anomaly is the negation of the amount of D$(p+4)$-brane charge in the vicinity of the probe. (This restriction to the vicinity of the probe is, of course, the heart of the matter since the anomaly cancels when we account for all of the D-branes and O-planes in spacetime.) Accounting for the charge of an O$(p+4)$-plane, we find that the anomaly is always $-N_f$ when the gauge group is $U(1)$ and $2^{p-1}-N_f$ when the gauge group is $SU(2)$. When the anomaly is negative, it signals that somewhere on the Coulomb branch there will be some singular behavior in the field theory that is not cured by quantum corrections. When $p>3$, this signals that we have left the domain of definition of the non-renormalizable effective field theory -- for instance, when $p=4$ a tadpole yields a negative kinetic energy term for the gauge field, destroying the consistency of the theory. We are, of course, still free to study the effective field theory away from this region, but we should not expect the field theory to have a strong-coupling fixed point. Instead, the UV completion of the theory involves extra degrees of freedom from string theory. When $p=4$ and the anomaly is positive, it turns out that there are strong-coupling fixed points; the inverse of the coupling constant parametrizes a relevant direction away from these points. When $p<3$, the theory is asymptotically free (or free in the case of $U(1)$ with $N_f=0$), and when the anomaly is negative we find singular behavior in the IR instead of in the UV. Specifically, singularities arise at points in the moduli space where degrees of freedom not captured by the moduli space become light. Note that this implies that taking the strong coupling limit and dimensional reduction do not commute, since in higher dimensions we found some new theories at strong coupling, some of which yield theories in lower dimensions with $E_n$ global symmetry, even though the strong coupling limit in lower dimensions only yields the theories in the $A_n$ and $D_n$ series \cite{danielsson:compact}. This non-commutativity will be important later in another context. The case $p=3$ interpolates between these discussions as a function of $N_f$. We emphasize that we are free to consider the effective field theories even when they are `anomalous' -- this anomaly is not, for example, an anomaly in a symmetry that we are trying to gauge, which would destroy the theory. However, if we are interested in global aspects of the anomalous non-renormalizable theories we need to complete them in the UV.

In this discussion, we have glossed over a crucial detail. Thus far, we only have enough D-branes to get a rank 16 gauge group, but we know that toroidal compactification of the heterotic string allows us to enhance the gauge symmetry. We also know that new gauge groups should be attainable because there is a point in moduli space with $E_8\times E_8$ gauge symmetry. The usual Narain compactification mechanism for gauge symmetry enhancement of heterotic strings cannot work here, since the winding number of type I strings is not conserved -- closed strings can break and end on the D9-branes. The resolution of this dilemma involves interesting non-perturbative physics. Although we are most interested in the case with a 2d probe, we will make a brief digression and review the other cases as well, as comparing and contrasting these different cases highlights a number of important features.

This problem was addressed in 9 dimensions in \cite{w:evidence}, where it was shown that in the supergravity approximation to the T-dual of type I on $S^1$ (a.k.a. type I$'$), the dilaton is forced to vary as a function of its position on $S^1/Z_2$, and points of enhanced gauge symmetry in the heterotic frame map to points in the type I$'$ moduli space where the string coupling diverges at an O8-plane. This suggests that D-branes, which become light near the O8-planes, might supply the necessary new gauge bosons. This is in fact correct: D0-brane bound states with an O8-plane can enhance the gauge symmetry of the type I$'$ theory\cite{kachru:extraD0s,lowe:extraD0s,nilles:bps,gaberdiel:extraD0s,green:enhance}. This modifies the O8-plane -- it becomes a new type of non-perturbative defect, namely an E8-plane plus a D8-brane. This D8-brane can be ejected away from the E8-plane, although the positions of the D8-branes in the theory are constrained by the requirement that the theory be strongly coupled near the E8-plane. This discussion was reproduced in field theory in \cite{s:5d,s:extremal,vafa:delPezzo}. For example, the dilaton profile of \cite{w:evidence} arises from a one-loop renormalization of the coupling constant, while the constraint on the D8-brane positions when there are extra D8-branes comes from the fact that the strong coupling fixed point does not have extra mass parameters.

Similar conclusions hold in lower dimensions: non-perturbative effects can cause O-planes to split. As expected from the heterotic frame, a compactification with $d$ compact dimensions can have a non-Abelian gauge group of arbitrary ADE-type with rank up to $16+d$ (in addition to Abelian factors). This O-plane splitting is particularly interesting in the case of D3-branes probing the 8-dimensional compactification, which is dual to F-theory on an elliptically fibered K3, since the Yang-Mills coupling is classically marginal in 4 dimensions, so the asymptotic freedom of the theory depends on the number of hypermultiplets in the theory. In particular, when too few D7-branes are near an O7-plane, the O7-plane splits and we discover the 24 D7-branes of the F-theory compactification \cite{sen:FOrientifolds,dasgupta:constant,s:fBranes}. Strong coupling fixed points exist for $SU(2)$ gauge theory with $1\le N_f\le 4$, corresponding to $N_f-1$ D7-branes at an H7-plane ($1\le N_f\le 3$)\cite{sw:argyresDouglas} or to four D7-branes at an O7-plane ($N_f=4$) \cite{sw}. We can also get fixed points describing 6, 7, or 8 D7-branes at an E7-plane by compactifying their higher-dimensional variants \cite{ganor:compact,s:branesFields}. As for the 5 dimensional probe, the (now complexified) dilaton profile of the F-theory compactification is reproduced by a coupling constant renormalization in the probe theory (although now this involves one-loop and instanton contributions). Indeed, in all of these theories -- including the non-Lagrangian ones \cite{minahan:E6,minahan:En} --  the low-energy behavior of the theory is encapsulated by an elliptic fibration over the Coulomb branch. The gapless low-energy theories correspond with the classification\cite{kodaira} of singular fibers in elliptic fibrations over a 1-(complex)-dimensional base, and the irreducible components of the singular fibers have an intersection matrix whose ADE type corresponds to that of the global symmetry on the probe \cite{s:branesFields}.

Moving down another dimension, we find the classical Coulomb branch $(\RR^3\times S^1)/Z_2$ when the gauge group is $SU(2)$ and $\RR^3\times S^1$ when the gauge group is $U(1)$; the fourth modulus arises in both cases from dualizing the photon. However, quantum corrections modify this result, so that we end up with an ALF space \cite{sw:3d}. In fact, the origin has a singularity of the same type that is developed by the K3 of the dual M-theory background -- that is, our D2-brane is probing this background, in a similar vein as the previous section\cite{s:K3}.  In particular, this means that at points in the moduli space of the M-theory compactification where the K3 is non-singular, quantum corrections eliminate the singularity at the origin. This corresponds to the absence of new light states at the origin: neither the gauge nor global symmetries are enhanced there. This absence of light states is quite surprising for $U(1)$ with $N_f=1$, as it implies that a D2-D6 string is massive even when the D-branes coincide. A similar surprise occurs for $SU(2)$ with $N_f=0,1$, as charged strings stretched across an O6-plane which coincides with zero or one D6-branes remain massive, even as their classical length vanishes.\footnote{The case of the D6-D6 string actually has a perturbative explanation: the orientifolding operation projects out the charged ground states of these strings, so that the lightest charged strings have mass of order $M_s$ from oscillator modes \cite{polchinski:consistent}. This is reflected in the fact that the D6-branes' gauge group is $SO(2)\cong U(1)$. This effect can also be corroborated at strong string coupling by lifting to M-theory, where an O6-plane which coincides with a D6-brane is described by the double cover $\bar \N$ of the Atiyah-Hitchin centered $SU(2)$ 2-monopole moduli space (which may itself be characterized as a monopole moduli space \cite{kapustin:singularMonopoles}). For, the charged D6-D6 string corresponds to an M2-brane wrapping the ``bolt'' 2-cycle of $\bar\N$, whose area is always positive when $g_s>0$ \cite{sen:ahBolt}.} Lifting to M-theory provides a natural explanation for the appearance of an ALF space, as the D6-brane becomes a KK-monopole \cite{sen:hetAn,sen:d6KK}. Furthermore, since the IR limit of the field theory corresponds to strong coupling, which decompactifies the M-theory circle, we expect that in the IR limit the field theory's moduli space should decompactify to an ALE space; indeed, this is the case -- the radius of the circle parametrized by the dual photon is proportional to the coupling constant $g_{3d}$ of the field theory.\footnote{We can show more explicitly that this decompactifying circle is the M-theory circle. First, we recall the relation $\vec r=\vec x M_s^2$ between physical positions $\vec x$ and Coulomb branch coordinates $\vec r$. The dimensionless coordinates that we hold fixed in the IR limit are, in turn, obtained via $\tilde{\vec r}=\vec r/g_{3d}^2=(\vec x M_s^2)/(g_s M_s)=\vec x/R_M$, where $R_M=g_s/M_s$ is the radius of the M-theory circle. In particular, one of these dimensionless coordinates is the dual photon; its period of $2\pi$ translates to a spacetime period of $2\pi R_M$, which clearly identifies this coordinate as the angular position around the M-theory circle.} In particular, the ALE manifolds obtained by decompactifying the $D_0$ and $D_1$ ALF manifolds in this way, which we will call the $D_0$ and $D_1$ ALE manifolds, are both the $A_1$ ALE manifold; similarly, the $D_2$ ALE manifold coincides with the $A_1$ ALE manifold, as the $D_2$ moduli space is not corrected from $(\RR^3\times S^1)/Z_2$, and as we decompactify this moduli space we zoom in at one of the $Z_2$ fixed points \cite{sw:3d,kapustin:dnQuivers}. Two more cases where the obvious correspondence between ALE and ALF manifolds breaks down are $U(1)$ with $N_f=0,1$, where the ALF manifolds are, respectively, the $A_{-1}$ $\RR^3\times S^1$ and $A_0$ Taub-NUT manifolds, both of which decompactify to $\RR^4$. Finally, using the exceptional isomorphism $D_3\cong A_3$, we find that the $D_3$ and $A_3$ ALE manifolds are the same, but it turns out that the corresponding ALF manifolds are different \cite{minerbe:rigidTaubNUT}.

Finally, we arrive in two dimensions. Classically, the Coulomb branch is $\RR^4/Z_2$ when the gauge group is $SU(2)$ and $\RR^4$ when the gauge group is $U(1)$, where the metrics on these spaces are flat, and again quantum corrections will modify this in order to be consistent with heterotic-type IIA duality. However, there is a catch: rather than develop an ADE singularity, the Coulomb branch develops an infinitely long throat with torsion.\footnote{As we mentioned earlier, the meaning of the moduli space is different in two dimensions compared to higher dimensions. One might fear that we are no longer allowed to ignore the Higgs branch in this case, as the low-energy sigma model we obtain from the gauge theory should have the whole moduli space as its target space. However, this infinite throat has the effect of decoupling the Higgs and Coulomb branches of our gauge theory in the IR\cite{w:comments,sk:decoupling,w:higgs}, as they are infinitely far away from each other, so they describe decoupled sigma models. More precisely, when neither branch of moduli space is lifted, these sigma models become singular \cite{aspinwall:theta,w:comments}; lifting one branch yields a non-singular sigma model on the other.

The singularity of the Higgs branch is rather interesting, as non-renormalization theorems protect the metric on the Higgs branch, so it seems strange for it to develop the throat that the decoupling arguments require. The solution involves a breakdown in the moduli space approximation near the origin of the Higgs branch. The throat appears once we employ the appropriate variables, which in this case happen to be the fields in the vector multiplets \cite{aharony:LSTmatrix,verlinde:higgsCoulomb,w:holoHiggs}.} These throats still have an ADE classification \cite{s:coulomb,s:ns5holo,kapustin:hlst} and are related (after we partially compactify their asymptotic regions, so that we have a circle to T-dualize) by mirror symmetry to the associated ALF spaces\cite{ooguri:ADE,w:comments,s:coulomb,brodie:2dMirror,sethi:mirror,kapustin:mirror,tong:mirror,harvey:monopoles3,jensen:doubledKK}.\footnote{This statement requires some quantum correction. The ALF space does not break the isometry around the circle, so this isometry does not yield a zero mode of the KK-monopole. This is in stark contrast with the throat, which does break its circle's isometry\cite{harvey:monopoles2,tong:mirror}. The resolution to this apparent zero mode mismatch is that the throat's circle is dual to a ``winding space," parametrized by the value of the B-field, in which the ALF singularity is localized\cite{sen:monopoles}. It is argued in \cite{harvey:monopoles,harvey:monopoles3,jensen:doubledKK,brodie:2dMirror} that there is a throat in the quantum-corrected ALF geometry at this point in winding space. In spite of the fact that the string-corrected ALF spaces and throats both have a throat, we will nevertheless follow historical convention and refer only to the latter as ``throats."} They describe the spacetimes of type IIB NS5-branes, as we should expect from S-duality which yields a description in which the D1-brane has become a fundamental string. In this S-dual picture, the O5-planes which carry negative D5-brane charge are replaced by 5-planes which carry negative NS5-brane charge \cite{kutasov:orbSolitons}. In fact, this S-dual frame allows us (at least, away from a point with enhanced gauge symmetry, where there are extra NS5-branes) to go full circle and return to type IIA on K3 \cite{vafa:dManifolds,kutasov:orbSolitons,Sen:1996na}, since a single T-duality takes us to (un-orientifolded) type IIA on $T^4/Z_2$. The positions of the NS5-branes are mapped, under this T-duality, to the blow-up modes of the $T^4/Z_2$ orbifold points. The S-dual frame is also interesting because it allows us to imagine the Coulomb branch of the HLST as resembling an actual field configuration in spacetime. In particular, this picture suggests that we should expect to find torsion throughout the Coulomb branch (and not only localized near singularities), since NS5-brane charges are not cancelled locally.\footnote{The existence of various localized objects allows us to evade supergravity theorems that forbid the existence of fluxes on compact spaces \cite{papadop:noGo,deWit:warp,maldacena:noGo,sk:noGo}.}

We can easily describe the metric and torsion in the $A_n$ and $D_n$ cases in the limit where the D5-branes coincide at the origin \cite{s:coulomb}. With the definition
\be\label{eq:k} k=\piecewise{N_f}{U(1)}{2(N_f-1)}{SU(2)}\ ,\ee
the metric and torsion are
\be\label{eq:nicer} ds^2 = \parens{\frac{1}{g_{2d}^2}+\frac{k}{r^2}}d\vec r^2\ ,\quad H=-k d\Omega^3\ ,\ee
where $d\Omega^3$ is the volume form on the unit 3-sphere centered at the origin. We now focus on the IR limit: $g_{2d}\to\infty$ with $\tilde{\vec r}=\vec r/g_{2d}$ finite (as well as $\tilde{\vec m}_i=\vec m_i/g_{2d}$, where $\vec m_i$ are the hypermultiplet masses) \cite{w:higgs,aharony:LSTmatrix};\footnote{\label{ft:orderG}This rescaling provides an intuitive explanation for why the origin of the Coulomb branch is infinitely far away: any finite value of $\tilde r$ corresponds to an infinite value of $r$, but the scalars $\vec r$ are the ones with the correct scaling for the Higgs branch \cite{aharony:LSTmatrix}. This rescaling also allows us to clearly see that the $g_{2d}\to\infty$ limit is the IR limit, since the potential on the Coulomb branch is, schematically, $\frac{1}{4g_{2d}^2}\Tr [r,r]^2=\frac{g_{2d}^2}{4}\Tr [\tilde r,\tilde r]^2$, so moving off the Coulomb branch costs an energy of order $g_{2d}$.} in these variables, we have
\be\label{eq:nicer2} ds^2 = \parens{1+\frac{k}{\tilde r^2}}d\tilde{\vec r}^2\ ,\quad H=-k d\Omega^3\ .\ee
These coordinates have the benefit of making it clear that the origin is at infinite distance from any other point. However, in order to analyze the throat in more detail we focus on the region with $\tilde r \ll \sqrt{k}$ and make the further change of coordinates $\phi = \sqrt{k}\log \sqrt{\tilde r^2/k}$. This yields
\be\label{eq:throat} ds^2 = d\phi^2 + kds_3^2\ ,\quad H=-k d\Omega^3\ ,\quad Q=\sqrt{2/k}\ ,\ee
where $ds_3^2$ is the metric on the unit 3-sphere and $Q$ is the background charge of a supersymmetrized linear dilaton field $\phi$ that diverges at the origin. We thus find a supersymmetric level $k$ $SU(2)$ WZW model times a supersymmetrized linear dilaton. (In truth, for the $D_n$ theories, we get an $SO(3)=SU(2)/Z_2$ WZW model.)

Actually, in the cases where $k<2$ ($U(1)$ or $SU(2)$ with $N_f=0,1$) there is no Higgs branch, so we do not expect a singularity in the Coulomb branch \cite{s:coulomb}. That is, the above description of the moduli space breaks down in these cases. (This is important, as the $U(1)$ with $N_f=0,1$ and $SU(2)$ with $N_f=0$ cases provide a local description of the smooth K3 surface from the previous section.) This breakdown in the moduli space has been argued in the $U(1)$ with $N_f=1$ case to arise from a bound state, called a quantum Higgs branch, that is localized at the origin \cite{w:higgs,brodie:2dMirror,aharony:LSTmatrix,harvey:quantumHiggs}. We emphasize that this breakdown when $N_f\le 1$ is expected to be indicative not of a sickness in the Coulomb branch CFT, but rather of a need to describe the region near the origin with different variables, as it would be strange for a singularity to be present when there is no classical Higgs branch emerging from the origin \cite{s:coulomb,aharony:LSTmatrix}. As we discuss below, the desired non-singular description of the moduli space is provided by string dualities. However, even without these we can handle the cases with $k=0$: they are described by \eqref{eq:nicer2}, which is $\RR^4$ for $U(1)$ with $N_f=0$ (as is clear from the fact that the theory is free) and $\RR^4/Z_2$ for $SU(2)$ with $N_f=1$.

Restricting to the cases with a throat, it turns out that they are in one-to-one correspondence with modular invariants of $\widehat{SU(2)}_{k-2}$. The ADE classification of these modular invariants \cite{ciz1,ciz2} yields an ADE classification of throats which corresponds precisely with the ADE classification of $(4,4)$ theories with a 4-dimensional Coulomb branch (and $N_f>1$ in the A and D cases) \cite{s:coulomb}.\footnote{Technically, for this correspondence to be perfect we need to include $A_{N_f-1}$ and $D_{N_f}$ throats for all $N_f\ge 2$, ignoring the exceptional isomorphisms $D_2\cong A_1\times A_1$ and $D_3\cong A_3$.} In these cases, the quantum Higgs branch causes no complications, as it is simply part of the classical Higgs branch \cite{w:higgs}.

Recalling that the quantum geometry of a K3 surface can have singularities of rank up to 20, we find the remarkable result that all 20 singularities are on the same footing in these throats, even though in the type IIA geometry some of the singularities arise from classical singularities while others arise from singularities in the quantum geometry. This uniformization will persist as our ADE systems increase in rank when we toroidally compactify our type IIA K3 background below. We thus reproduce, in a brane language, one of the main features of the Narain compactification mechanism for heterotic gauge symmetry enhancement.

\begin{figure}
\includegraphics[width=\textwidth]{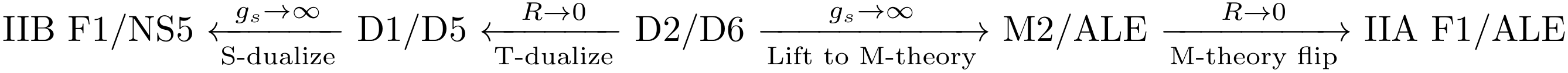}
\caption{Duality between NS5-brane and KK-monopole}\label{fig:nskkmirror}
\end{figure}

This difference between the 2d case, where we have throats, and the 3d case, which has singularities, arises from the fact that when we compactify the 3d theory with coupling constant $g_{3d}$ on a circle of radius $R$, the one-loop corrections to the Coulomb branch depend on the dimensionless quantity $\gamma=g_{3d}^2R$ \cite{s:coulomb,s:ns5}. When $\gamma$ is large, we are in the 3d regime where the Higgs and Coulomb branches intersect and the origin of the Coulomb branch has an ADE singularity. In contrast, when $\gamma$ is small, the Coulomb branch develops an infinite throat that decouples the Higgs and Coulomb branches\cite{w:comments,sk:decoupling,w:higgs}. Therefore, the 2d ($R\to0$) and IR ($g_{3d}\to\infty$) limits do not commute, as taking either limit eliminates the theory's dependence on $\gamma$. Figure \ref{fig:nskkmirror} illustrates how $\gamma$ interpolates between the physics of the type IIA KK-monopole and the type IIB NS5-brane.

In fact, by combining this reasoning with 3d mirror symmetry we can demonstrate the claim we made above that $A_n$ throats and ALF spaces are related by 2d mirror symmetry \cite{s:coulomb,brodie:2dMirror,sethi:mirror}. If we use 3d mirror symmetry\cite{s:3dmirror}\footnote{Actually, as emphasized by \cite{aganagic:gaugeMirror,tong:mirror2d}, this argument requires us to use the version of 3d mirror symmetry that is valid away from the infrared \cite{kapustin:mirror3}, since otherwise we are restricted to studying both sides of the duality at $\gamma=\infty$. \cite{aganagic:gaugeMirror,tong:mirror2d} also demonstrated that these same arguments can allow one to derive 2d mirror pairs with only $(2,2)$ supersymmetry. This is particularly interesting because these theories can have compact target spaces.} before taking $R\to0$ then we obtain a theory whose Higgs branch is identical to the original theory's Coulomb branch. The moduli space is then determined by a hyper-K\"ahler quotient, so it is the same in 2d or 3d.\footnote{Worldsheet instantons modify the physics of the 2d ALF gauged linear sigma model, but the standard description of the NLSM is unable to account for this \cite{harvey:monopoles3,jensen:doubledKK}.} However, the Higgs branch coordinate mirror to the dual photon parametrizes a circle with radius proportional to $\sqrt{\gamma}$.\footnote{$\gamma$ can affect the Higgs branch in the mirror theory because it is the coupling constant for a \emph{twisted} vector multiplet \cite{kapustin:mirror3}.} More precisely, in the $R\to 0$ limit we obtain a 2d NLSM whose target space is the appropriate ALF manifold whose asymptotic region at infinity is locally $\RR^3\times S^1$ where the radius of the $S^1$ behaves as stated. That compactification of the original theory's Coulomb branch yields a T-dual NLSM can then be understood by noting that the radius of the Wilson line around the compactifying direction (which is one of the coordinates on the Coulomb branch of the 2d theory) is proportional to $1/\sqrt{\gamma}$.\footnote{See \cite{harvey:monopoles2,s:coulomb} for the metric of this compactified theory's Coulomb branch, which interpolates between the 2d and 3d limits \cite{s:coulomb,brodie:2dMirror}.}

As promised, we can now employ dualities to describe the $N_f=0,1$ cases where the throat description breaks down. For $U(1)$ with $N_f=1$, we use the T-duality between a single NS5-brane and an $A_0$ ALF space (i.e. a Taub-NUT space) to confirm the absence of a singularity. When $\gamma=\infty$, this agrees with the familiar 3d IR mirror symmetry relation between $U(1)$ with $N_f=1$ and a free hypermultiplet (whose expectation values parametrize $\RR^4$); decreasing $\gamma$ compactifies the Taub-NUT circle. Similarly, 3d IR mirror symmetry maps pure $U(1)$ to a free hypermultiplet, but now the moduli space for finite $\gamma$ is $\RR^3\times S^1$. The 3d mirror symmetry argument breaks down for the $SU(2)$ theories with $N_f=0,1$, but the reasoning illustrated in figure \ref{fig:nskkmirror} still holds. This is particularly clear for $N_f=1$, as the S-dual of the D5+O5 configuration is T-dual to IIA on $\RR^4/Z_2$.

\bibliographystyle{utphys}

\bibliography{Refs}

\end{document}